\documentclass[lettersize,journal]{IEEEtran}
\IEEEoverridecommandlockouts
\usepackage[utf8]{inputenc}
\usepackage{multirow}
\usepackage{booktabs}
\usepackage{balance}
\usepackage[hidelinks]{hyperref}
\usepackage{comment}
\usepackage{enumitem}
\usepackage{ragged2e}
\usepackage{url}
\usepackage{stfloats}
\usepackage{tikz}
\usepackage[nobreak,space,compress]{cite}

\def\BibTeX{{\rm B\kern-.05em{\sc i\kern-.025em b}\kern-.08em T\kern-.1667em\lower.7ex\hbox{E}\kern-.125emX}}

\usetikzlibrary{shapes.geometric, positioning, arrows.meta, calc, backgrounds}

\newcommand{\approach}{AgenTag}
\newcommand{\dataset}{AIDev}
\newcommand{\nagentprs}{\NAgentPRs}
\newcommand{\nhumanprs}{\NHumanPRs}

\newcommand{\NAgentPRs}{33,580}

\newcommand{\NHumanPRs}{6,618}
\newcommand{\NHumanWithCommits}{6,403}
\newcommand{\PctHumanWithCommits}{97}

\newcommand{\BalancedCap}{2,500}
\newcommand{\NCodexPRs}{21,793}
\newcommand{\PctCodexPRs}{64.9}
\newcommand{\NCopilotPRs}{4,967}
\newcommand{\PctCopilotPRs}{14.8}
\newcommand{\NDevinPRs}{4,822}
\newcommand{\PctDevinPRs}{14.4}
\newcommand{\NCursorPRs}{1,540}
\newcommand{\PctCursorPRs}{4.6}
\newcommand{\NClaudePRs}{458}
\newcommand{\PctClaudePRs}{1.4}
\newcommand{\CursorCommitterPct}{92.5}

\newcommand{\ClaudeCoauthorPctOther}{0.2}
\newcommand{\CursorPRsWithClaudeTrailer}{11}
\newcommand{\NFeatTotal}{53}
\newcommand{\NFeatKept}{41}

\newcommand{\PRTextProbe}{0.766}
\newcommand{\PRTextContr}{0.825}

\newcommand{\CommitProbe}{0.765}
\newcommand{\CommitContr}{0.771}

\newcommand{\CodeProbe}{0.558}
\newcommand{\CodeContr}{0.589}

\newcommand{\BehavioralProbe}{0.832}
\newcommand{\BehavioralContr}{0.847}

\newcommand{\AllTextProbe}{0.851}
\newcommand{\AllTextContr}{0.890}
\newcommand{\AllTextContrTwo}{0.89}
\newcommand{\FusionProbe}{0.878}
\newcommand{\FusionContr}{0.909}

\newcommand{\FusionCodeProbe}{0.885}
\newcommand{\FusionCodeContr}{0.907}

\newcommand{\WFBalanced}{0.909}
\newcommand{\WFBalancedTwo}{0.91}
\newcommand{\MFBalanced}{0.857}

\newcommand{\CodexBalanced}{0.955}

\newcommand{\CopilotBalanced}{0.975}

\newcommand{\DevinBalanced}{0.886}

\newcommand{\CursorBalanced}{0.853}

\newcommand{\ClaudeBalanced}{0.616}

\newcommand{\WFNatural}{0.957}
\newcommand{\WFNaturalTwo}{0.96}
\newcommand{\MFNatural}{0.840}
\newcommand{\MFNaturalTwo}{0.84}
\newcommand{\CodexNatural}{0.984}
\newcommand{\CodexNaturalTwo}{0.98}
\newcommand{\CopilotNatural}{0.983}
\newcommand{\CopilotNaturalTwo}{0.98}
\newcommand{\DevinNatural}{0.899}
\newcommand{\DevinNaturalTwo}{0.90}
\newcommand{\CursorNatural}{0.779}
\newcommand{\CursorNaturalTwo}{0.78}
\newcommand{\ClaudeNatural}{0.556}
\newcommand{\ClaudeNaturalTwo}{0.56}
\newcommand{\OursNatW}{0.957}
\newcommand{\XgbNatW}{0.954}
\newcommand{\PNatW}{0.036}
\newcommand{\DeltaNatW}{-0.55}
\newcommand{\OursNatM}{0.840}
\newcommand{\XgbNatM}{0.854}
\newcommand{\PNatM}{0.002}
\newcommand{\DeltaNatM}{+0.78}

\newcommand{\PBalW}{0.821}

\newcommand{\MaxBaselineDiff}{0.014}
\newcommand{\CodeMiniLM}{0.59}
\newcommand{\CodeCodeBERT}{0.56}
\newcommand{\CodeGraphCodeBERT}{0.59}
\newcommand{\Codehandcraftedcontent}{0.34}
\newcommand{\AllTextNoCode}{0.890}
\newcommand{\AllTextPlusCode}{0.892}
\newcommand{\StdWF}{0.947}
\newcommand{\StdMF}{0.805}
\newcommand{\RepoWF}{0.900}
\newcommand{\RepoMF}{0.738}

\newcommand{\LangMFMean}{0.626}
\newcommand{\LangMFMedian}{0.670}
\newcommand{\LangMFLo}{0.433}
\newcommand{\LangMFHi}{0.713}
\newcommand{\TimeWF}{0.933}
\newcommand{\TimeMF}{0.749}
\newcommand{\TimeBaseWF}{0.947}
\newcommand{\TimeBaseMF}{0.805}

\newcommand{\TimeDevinTwo}{0.71}

\newcommand{\TimeCursorTwo}{0.65}
\newcommand{\TimeClaude}{0.433}
\newcommand{\TimeClaudeTwo}{0.43}
\newcommand{\DegenNTest}{8,913}
\newcommand{\DegenMF}{0.199}
\newcommand{\DegenWF}{0.997}
\newcommand{\DCodexBody}{-0.75}
\newcommand{\DCodexTitle}{-0.73}
\newcommand{\DCodexCommits}{-0.67}
\newcommand{\DCodexCommitLen}{-0.62}

\newcommand{\DCopilotBody}{+0.85}

\newcommand{\DCopilotGini}{-0.71}

\newcommand{\DDevinCommitLen}{+0.71}

\newcommand{\DDevinConv}{+0.31}

\newcommand{\DCursorBullets}{-0.52}

\newcommand{\DClaudeCommitLen}{+0.81}
\newcommand{\DClaudeMultiline}{+0.81}

\newcommand{\DClaudeComments}{+0.39}
\newcommand{\CopilotInitPlan}{97}
\newcommand{\CopilotLinksIssue}{84}

\newcommand{\CopilotSingleLine}{97}

\newcommand{\ClaudeMedCommitChars}{231}
\newcommand{\ClaudeSummaryBullets}{74}

\newcommand{\CursorMedCommitChars}{58}
\newcommand{\CursorHeaders}{28}
\newcommand{\DevinConventional}{46}

\newcommand{\CodexMedCommitChars}{35}
\newcommand{\MultilineCopilotRaw}{0.55}
\newcommand{\MultilineCopilotStripped}{0.02}
\newcommand{\MultilineDevinRaw}{0.84}
\newcommand{\MultilineDevinStripped}{0.28}

\newcommand{\LadderLatentWF}{0.947}
\newcommand{\LadderLatentWFTwo}{0.95}
\newcommand{\LadderLatentMF}{0.803}
\newcommand{\LadderLatentCodex}{0.983}

\newcommand{\LadderLatentCopilot}{0.971}

\newcommand{\LadderLatentDevin}{0.878}

\newcommand{\LadderLatentCursor}{0.739}

\newcommand{\LadderLatentClaude}{0.446}
\newcommand{\LadderLatentClaudeTwo}{0.45}
\newcommand{\LadderNamesWF}{0.951}

\newcommand{\LadderNamesMF}{0.815}
\newcommand{\LadderNamesCodex}{0.984}

\newcommand{\LadderNamesCopilot}{0.973}

\newcommand{\LadderNamesDevin}{0.883}

\newcommand{\LadderNamesCursor}{0.766}

\newcommand{\LadderNamesClaude}{0.467}

\newcommand{\LadderBoilerWF}{0.961}

\newcommand{\LadderBoilerMF}{0.837}
\newcommand{\LadderBoilerCodex}{0.988}

\newcommand{\LadderBoilerCopilot}{0.975}

\newcommand{\LadderBoilerDevin}{0.919}

\newcommand{\LadderBoilerCursor}{0.800}

\newcommand{\LadderBoilerClaude}{0.503}

\newcommand{\LadderRawWF}{0.973}
\newcommand{\LadderRawWFTwo}{0.97}
\newcommand{\LadderRawMF}{0.894}
\newcommand{\LadderRawCodex}{0.992}

\newcommand{\LadderRawCopilot}{0.985}

\newcommand{\LadderRawDevin}{0.945}

\newcommand{\LadderRawCursor}{0.832}

\newcommand{\LadderRawClaude}{0.717}
\newcommand{\LadderRawClaudeTwo}{0.72}
\newcommand{\DiscloseTotalWF}{+0.026}
\newcommand{\DiscloseTotalWFPlain}{0.026}
\newcommand{\NamesDeltaWF}{0.004}

\newcommand{\BoilerDeltaWF}{+0.010}
\newcommand{\TrailerDeltaWF}{+0.012}
\newcommand{\MarkerCostCodex}{0.009}

\newcommand{\MarkerCostCopilot}{0.014}

\newcommand{\MarkerCostClaude}{0.271}

\newcommand{\TrailerCostMaxMajor}{0.010}

\newcommand{\CopilotNameRateRaw}{98}
\newcommand{\CopilotNameRateProse}{20}

\newcommand{\DevinNameRateRaw}{98}
\newcommand{\DevinNameRateProse}{19}

\newcommand{\CodexNameRateRaw}{83}
\newcommand{\CodexNameRateProse}{7}

\newcommand{\NameRegexProse}{0.652}
\newcommand{\NameRegexProseTwo}{0.65}
\newcommand{\NameRegexFloor}{0.517}

\newcommand{\NameRegexGainSigned}{+0.135}
\newcommand{\MaxNameResidual}{0.9}
\newcommand{\ClaudeNameResidual}{5.9}
\newcommand{\FeatAllWF}{0.883}
\newcommand{\FeatLatentWF}{0.843}
\newcommand{\FeatStrictWF}{0.762}
\newcommand{\FeatMarkersWF}{0.788}
\newcommand{\HvAPRText}{0.85}

\newcommand{\HvACode}{0.65}

\newcommand{\HvAAllText}{0.88}

\newcommand{\HvAFusion}{0.89}

\newcommand{\HvAStarMatchedFusion}{0.86}

\newcommand{\HvAWithinRepo}{0.81}
\newcommand{\NSharedRepos}{810}
\newcommand{\ContamFloorN}{334}
\newcommand{\ContamFloorPct}{5.0}
\newcommand{\ContamFloorPctZero}{5}
\newcommand{\GenericTrailerPct}{14.7}
\newcommand{\ContamFiveN}{693}
\newcommand{\ContamFivePct}{10.5}
\newcommand{\ContamFiveMarker}{18.8}

\newcommand{\ContamSevenMarker}{26.0}

\newcommand{\ContamNineN}{87}
\newcommand{\ContamNinePct}{1.3}
\newcommand{\ContamNineMarker}{39.1}
\newcommand{\ContamNineEnrich}{7.7}
\newcommand{\AuditN}{40}
\newcommand{\AuditDisclosed}{14}
\newcommand{\AuditTemplate}{30}
\newcommand{\AuditInconclusive}{10}
\newcommand{\AuditPMin}{0.94}
\newcommand{\OpenPRTextL}{0.757}
\newcommand{\EnrollPRTextL}{0.573}
\newcommand{\OpenPRTextN}{0.739}
\newcommand{\EnrollPRTextN}{0.496}
\newcommand{\OpenCommitL}{0.755}
\newcommand{\EnrollCommitL}{0.526}
\newcommand{\OpenCommitN}{0.704}
\newcommand{\EnrollCommitN}{0.504}
\newcommand{\OpenAllTextL}{0.824}
\newcommand{\EnrollAllTextL}{0.643}
\newcommand{\OpenAllTextN}{0.750}
\newcommand{\EnrollAllTextN}{0.560}
\newcommand{\OpenFusionL}{0.837}
\newcommand{\EnrollFusionL}{0.674}
\newcommand{\OpenFusionN}{0.756}
\newcommand{\EnrollFusionN}{0.595}
\newcommand{\OpenFusionLTwo}{0.84}

\newcommand{\OpenGain}{+0.081}
\newcommand{\OpenGainPRText}{+0.018}
\newcommand{\EnrollGain}{+0.079}
\newcommand{\NewAgentL}{0.551}
\newcommand{\NewAgentN}{0.612}
\newcommand{\OpenMedian}{0.816}
\newcommand{\OpenDevin}{0.906}
\newcommand{\OpenCursor}{0.864}
\newcommand{\OpenRestLo}{0.796}
\newcommand{\OpenRestHi}{0.816}
\newcommand{\TradeoffCommitGain}{+0.065}
\newcommand{\TradeoffPatchGain}{+0.019}

\newcommand{\RepoMFMedian}{0.729}

\newcommand{\LangWFLo}{0.80}
\newcommand{\LangWFHi}{0.98}

\newcommand{\WFNaturalMedian}{0.957}
\newcommand{\LadderLatentWFMedian}{0.947}
\newcommand{\HvAFusionMedian}{0.886}
\newcommand{\MeanMedianMaxGap}{0.003}
\newcommand{\SeedPMacroLo}{0.02}
\newcommand{\SeedPMacroHi}{0.15}
\newcommand{\SeedClaudeSpread}{0.012}
\newcommand{\SeedRepoMeanSpread}{0.011}
\newcommand{\SeedLoloSpreadHi}{0.070}
\newcommand{\SeedLoloSpreadLo}{0.016}
\newcommand{\SeedLoloBigLang}{HTML}

\newcommand{\SeedLoloSmallLang}{Python}

\newcommand{\SeedRepoBandLo}{0.70}
\newcommand{\SeedRepoBandHi}{0.78}

\usepackage{tcolorbox}
\tcbset{
    rqbox/.style={
        colback=black!5!white, 
        boxrule=0pt,          
        colframe=black!5!white,
        sharp corners,        
        boxsep=5pt,           
        left=0pt, right=0pt, top=0pt, bottom=0pt, 
        fontupper=\normalsize 
    }
}

\begin{document}

\title{\approach: Attribution of AI Coding Agents from Behavioral Fingerprints}

\author{
  Taher A. Ghaleb~\href{https://orcid.org/0000-0001-9336-7298}{\includegraphics[scale=0.06]{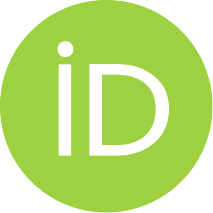}}

  \thanks{Taher A. Ghaleb is with the Department of Computer Science, Trent University, Peterborough, Ontario, Canada (e-mail: taherghaleb@trentu.ca).}
}

\markboth{}
{Ghaleb: \approach: Attribution of AI Coding Agents from Behavioral Fingerprints}

\maketitle

\begin{abstract}
  AI coding agents increasingly author pull requests (PRs), often under developers' own accounts, obscuring who actually produced a change. Reliable attribution is important for repository governance, empirical studies of AI-assisted software development, and measuring the impact of AI coding agents. Existing work focuses on closed-set identification of known agents, leaving the practical limits of open-world AI coding agent attribution largely unexplored. In this paper, we present \approach{}, a multimodal framework for open-world AI coding agent attribution, evaluated on \dataset{}, comprising \nagentprs{} PRs from five AI coding agents and \nhumanprs{} human-authored PRs. We represent each PR using textual, behavioral, and code-based modalities, and compare conventional classification with supervised contrastive learning for open-set recognition and few-shot enrollment of previously unseen agents. \approach{} identifies authoring agents with a weighted F1 of $\WFNaturalTwo$ (macro F1 of $\MFNaturalTwo$), distinguishes AI- from human-authored PRs with a balanced F1 of $\HvAFusion$, and detects previously unseen agents with an AUC of $\OpenFusionLTwo$. We further show that PR descriptions and commit messages provide nearly all of the attribution signal, whereas code diffs contribute little across multiple representations, indicating that coding agents are distinguished primarily by how they communicate changes rather than by the code they generate. Moreover, these behavioral fingerprints persist after removing explicit self-disclosed markers, demonstrating that attribution relies largely on latent stylistic characteristics. These findings show that reliable attribution of AI coding agents is feasible and clarify the practical trade-offs between attribution accuracy and the information required to achieve it.
\end{abstract}

\begin{IEEEkeywords}AI coding agents, Code authorship attribution, Contrastive learning, Open-set recognition
\end{IEEEkeywords}

\section{Introduction}
\label{sec:intro}
\IEEEPARstart{T}{he} widespread adoption of AI coding agents has introduced a new class of contributors to GitHub repositories, as they increasingly author pull requests (PRs) that implement features, fix bugs, and update documentation~\cite{li2025aidev,robbes2026agentic}.
Though PRs opened by autonomous agent accounts are explicitly labeled, the same agents are also invoked locally, with the resulting PRs submitted under developers' own accounts~\cite{li2025aidev}. This makes the submitting account an unreliable indicator of whether a change was authored by a human or an AI coding agent, or, if AI-generated, which agent produced it.
This disconnect between submitter identity and authorship has practical consequences.
Repository maintainers cannot reliably enforce disclosure policies or tool restrictions~\cite{github_tos}.
Datasets that infer authorship from submitter identity risk mislabeling AI-authored PRs as human-authored~\cite{li2025aidev}, threatening the validity of downstream empirical studies~\cite{ghaleb2019noise}.
Accurately measuring the impact of AI coding agents on software development likewise requires reliable attribution~\cite{barke2023grounded}.

Recent work has shown that AI-generated software artifacts contain detectable authorship signals.
Closed-set attribution among known AI coding agents, using PR structure, commit messages, and code changes, achieves high accuracy~\cite{ghaleb2026fingerprinting}, while code stylometry has long been used for human authorship attribution~\cite{caliskan2015anonymizing,burrows2007source,abuhamad2018large}.
However, four challenges remain.
First, the ability to distinguish AI-authored from human-authored PRs has not been established.
Second, it is unclear whether attribution relies on latent behavioral fingerprints or on explicit markers such as PR templates and ``Generated with'' commit trailers.
Third, existing approaches assume a closed set of known agents and cannot detect unseen ones.
Finally, the software artifacts needed for accurate attribution remain unknown, especially because commit messages and code diffs require extra collection, whereas PR text is readily available.

In this paper, we propose \approach{}, which frames agent attribution as a multimodal, learned, open-world problem rather than a closed-set classification task.
\approach{} represents a PR through four complementary streams (its title and description, commit messages, code diffs, and behavioral features) and learns a metric space with supervised contrastive learning~\cite{khosla2020supcon} in which PRs from the same agent cluster together.
This representation enables identifying and recognizing unseen AI coding agents and enrolling a new agent from a few labeled examples without retraining, capabilities a closed-set classifier lacks.
To identify agents by latent behavior rather than self-disclosed artifacts, we strip self-disclosed identity from all representations by default in three tiers (commit trailers, tool URLs, vendor template, and residual self-naming) and separately measure what each tier contributes.

We evaluate \approach{} on the \dataset{} dataset~\cite{li2025aidev}, which after cleaning contains \nagentprs{} PRs from five agents (OpenAI Codex, GitHub Copilot, Devin, Cursor, and Claude Code) and \nhumanprs{} human-authored PRs.
We embed the PR text and commit messages with a sentence encoder~\cite{reimers2019sbert} and the code diffs with a code-search encoder in the style of CodeBERT~\cite{feng2020codebert}, we attribute SHAP-based importance~\cite{lundberg2017shap} and Cliff's delta effect sizes~\cite{romano2006cliff} to make the behavioral characterization interpretable, and we report both weighted and macro F1 under the natural, imbalanced class distribution so that minority agents remain visible.

\smallskip\noindent\textbf{Contributions:}
Our contributions are as follows.
\vspace{-2pt}
\begin{itemize}[leftmargin=12pt, itemsep=1pt]
    \item We propose \approach{}, a learned metric space that identifies known agents, rejects agents it never saw, and enrolls a new one from ten labeled PRs without retraining, and we release a replication package~\cite{replication_package}.
    \item We show that agents are identified by how they write, not by what they write: PR and commit text nearly match the full multimodal model while code diffs add almost nothing across four code representations, inverting the ordering established for human authorship attribution.
    \item We show that the fingerprint reflects latent behavior, not a self-applied watermark: stripping trailers, names, tool URLs, and templates costs only $\DiscloseTotalWFPlain$ weighted F1, and the names alone cost $\NamesDeltaWF$ even though a regex over them achieves $\NameRegexProseTwo$ on the same text.
    \item We show that undisclosed AI authorship is detectable from PR text alone (balanced F1 $\HvAPRText$, or $\HvAFusion$ across modalities), and that at least $\ContamFloorPctZero\%$ of \dataset{}'s human-labeled PRs are agent-assisted by their own disclosure.
    \item We identify two measurement choices that invert conclusions here: markers must be stripped before feature extraction, not only before embedding, and an enrolled agent must be scored together with those already known.
\end{itemize}

\smallskip\noindent\textbf{Paper Organization:} 
The rest of this paper is organized as follows.
Section~\ref{sec:background} provides background on AI coding agents, contrastive representation learning, and open-set recognition.
Section~\ref{sec:approach} presents \approach{}, comprising the four signal modalities, their vector representations, and the contrastive metric space.
Section~\ref{sec:design} describes the study design, including the dataset, the evaluation metrics, and the statistical protocol.
Section~\ref{sec:results} reports results for the five research questions.
Section~\ref{sec:discussion} discusses implications and trade-offs.
Section~\ref{sec:threats} discusses validity threats.
Section~\ref{sec:related} reviews related work.
Section~\ref{sec:conclusion} concludes the paper and suggests future work.

\vspace{-5pt}
\section{Background}
\vspace{-1pt}
\label{sec:background}

\noindent\textbf{AI coding agents.}
AI coding agents are LLM-based tools that autonomously perform software engineering tasks, including editing code, running tests, and submitting pull requests (PRs) from natural-language instructions~\cite{li2025aidev}. Unlike code-completion assistants that provide in-editor suggestions~\cite{barke2023grounded}, coding agents work over entire repositories and generate complete PRs.
Li et al.~\cite{li2025aidev} identify five coding agents that account for most agentic PRs on GitHub: OpenAI Codex, GitHub Copilot, Devin, Cursor, and Claude Code. These agents vary in PR acceptance rates by task type~\cite{pinna2026comparing} and in how they modify and describe code~\cite{ogenrwot2026how,pham2026agentic}, suggesting that agent identity may be reflected in software artifacts.
Two deployment modes are relevant to attribution. In \emph{autonomous} mode, agents submit PRs through dedicated bot accounts, making authorship explicit. In \emph{human-mediated} mode, developers run agents locally and submit PRs under their own accounts, obscuring the true author. This mode is not hypothetical: \dataset{} can identify Claude Code from its commit trailer, since it commits under the developer's account~\cite{li2025aidev}, and we later find that at least $\ContamFloorPctZero\%$ of \dataset{}'s human-labeled PRs are in fact agent-assisted (Section~\ref{sec:rq4}). Human-mediated submission thus motivates AI-versus-human attribution and risks mislabeled datasets when identity relies solely on the submitting account.

\smallskip\noindent\textbf{Contrastive representation learning.}
Supervised contrastive learning learns an embedding in which examples from the same class are close together while examples from different classes are separated~\cite{khosla2020supcon}. Each class is represented by a prototype, computed as the centroid of its embeddings~\cite{snell2017prototypical}. Classification is performed by nearest-prototype search, while embedding similarity naturally supports verification of whether two inputs share the same class.
Prototype-based embeddings are widely used for open-world identity problems such as face recognition~\cite{schroff2015facenet} and speaker verification~\cite{snyder2018xvectors}, where new identities must be detected and enrolled without retraining.

\smallskip\noindent\textbf{Open-Set recognition and few-shot learning}
Closed-set classifiers assume that every test example belongs to one of the classes observed during training. Open-set recognition relaxes this assumption by allowing inputs that do not match any known class to be rejected as unknown~\cite{scheirer2013openset}. In prototype-based embeddings, this is achieved by thresholding the similarity between an input and the nearest prototype.
Few-shot learning complements open-set recognition by enabling new classes to be enrolled from only a few labeled examples~\cite{snell2017prototypical}. Each new class is represented by the centroid of its example embeddings, eliminating the need to retrain the embedding model. Together, these capabilities enable attribution in an open-world setting where new AI coding agents continually emerge.

\vspace{-5pt}
\section{Proposed Approach: \approach}
\label{sec:approach}
\approach{} attributes each PR to its authoring agent or to a human by turning four modalities into a metric space where PRs from the same author cluster, and then evaluates this space for identification, characterization, robustness to self-disclosed markers, AI-versus-human detection, and recognition and enrollment of unseen agents.

\vspace{-3pt}
\subsection{Signal Modalities}
We represent each PR through four modalities that differ in what they capture and in how readily they can be obtained.

\smallskip\noindent\textbf{1) PR text.} The PR title and description capture how a change is summarized and explain each change. This stream comes with every PR, including human-authored ones.

\smallskip\noindent\textbf{2) Commit messages.} Concatenated PR commit messages capture version-control communication style, such as conventional-commit prefixes~\cite{conventional_commits} and multi-paragraph bodies, and therefore require access to the PR's commits.

\medskip\noindent\textbf{3) Code diffs.} The lines added by a PR, extracted from its patches, capture code content and formatting. This stream requires file-level diffs and has the largest payload.

\begin{table*}[ht]
    \centering
    \vspace{-7pt}
    \caption{The 53 behavioral features by category ($41$ after correlation \& redundancy analysis). Features with $^{\dagger}$ are the eight explicit markers that RQ3 excludes to isolate latent signals.}
    \vspace{-9pt}
    \label{tab:features}
    \resizebox{\textwidth}{!}{
    \begin{tabular}{@{}p{2.6cm}p{16.1cm}@{}}
    \toprule
    \textbf{Category} & \textbf{Features} \\
    \midrule
    Commit patterns (8) &
      num\_commits;\quad avg/min/max/std commit length (characters);\quad
      multiline\_commit\_ratio$^{\dagger}$;\quad capitalized\_commit\_ratio$^{\dagger}$;\quad
      conventional\_commit\_ratio$^{\dagger}$ (Conventional-Commits prefix~\cite{conventional_commits}) \\
    \addlinespace
    PR structure (9) &
      title\_length, body\_length;\quad title\_word\_count, body\_word\_count;\quad
      has\_checklist$^{\dagger}$;\quad
      num\_links$^{\dagger}$;\quad num\_bullets$^{\dagger}$;\quad conventional\_title$^{\dagger}$;\quad num\_code\_blocks$^{\dagger}$ (fenced blocks) \\
    \addlinespace
    Code changes (19) &
      num\_files;\quad num\_files added/modified/removed/renamed;\quad
      num\_unique\_extensions, extension\_diversity;\quad test/config/doc file ratio;\quad
      avg/max directory depth;\quad total additions/deletions/changes;\quad add\_del\_ratio;\quad
      avg additions/deletions per file;\quad change\_concentration\_gini (Gini of per-file change sizes) \\
    \addlinespace
    Patch-level code (13) &
      num\_added/removed\_lines;\quad comment\_density, import\_density (per added line);\quad
      avg/max/std line length;\quad avg\_indentation;\quad trailing\_whitespace\_ratio;\quad
      num\_functions/classes\_added;\quad num\_conditionals, num\_loops (control-flow keywords) \\
    \addlinespace
    Temporal (4) &
      hour\_of\_day;\quad day\_of\_week;\quad is\_weekend;\quad is\_business\_hours (9-to-17) \\
    \bottomrule
    \end{tabular}
    }
\vspace{-8pt}
\end{table*}

\medskip\noindent\textbf{4) Behavioral features.}
We compute \NFeatTotal{} features that summarize commit patterns, PR structure, code-change shape, patch-level characteristics, and temporal behavior (Table~\ref{tab:features}). To reduce multicollinearity, we apply correlation-based hierarchical clustering at $|\rho|\geq0.70$~\cite{guyon2003introduction,dormann2013collinearity} followed by an $R^2$ redundancy filter~\cite{harrell2015regression,kuhn2013applied}, yielding \NFeatKept{} features.
Self-disclosed identity is handled at two levels. At the text level, we strip it in three cumulative tiers, since an agent discloses itself in more than one way and stripping only the most obvious form leaves the rest in place. Tier~1 strips structured trailers (e.g., ``Co-authored-by'', ``Signed-off-by'', ``Generated with''). Tier~2 strips tool URLs and vendor template, such as session links (\texttt{app.devin.ai}, \texttt{gh.io/copilot}) and template headers (``\texttt{\#\# Link to Devin run}'', ``\texttt{<!-- START COPILOT CODING AGENT TIPS -->}''), which exist only to announce the tool. Tiers~1 and~2 delete the whole line, since the line has no other content. Tier~3 strips residual mentions of a tool's own name that survive in prose, deleting the name token but keeping the surrounding sentence, because the writing style around it is precisely the latent signal we want to measure. Tier~3 is deliberately aggressive and also strips legitimate uses, such as a database ``cursor'', which biases the latent estimate downward and keeps it a conservative lower bound.
Stripping is applied before both embedding and feature extraction, so text-length and commit-format features are computed on latent text, which matters because a marker left in place is silently counted as behavior (RQ2). Our study uses the latent representation by default. RQ3 restores the tiers one at a time to quantify what each contributes.
Then, we distinguish eight stylistic features, such as template checkboxes, bullet and link structure, and commit-format conventions, from the remaining behavioral features. RQ3 excludes these stylistic features to assess whether attribution is driven by a latent behavioral fingerprint rather than explicit stylistic conventions.

\vspace{-7pt}
\subsection{Representations}
We encode PR text and commit messages with the Sentence-BERT encoder~\cite{reimers2019sbert} (all-MiniLM-L6-v2), which produces 384-dimensional embeddings per stream. We encode the title and the description separately and concatenate their embeddings, so the PR-text stream is 768-dimensional.
We encode code diffs using a Transformer code encoder over the added lines, mean-pooling token embeddings from up to the first $4,000$ characters of added code without task-specific fine-tuning. We compare a general sentence encoder, a code-search encoder in the style of CodeBERT~\cite{feng2020codebert}, and a structure-aware encoder pretrained on data-flow, GraphCodeBERT~\cite{guo2021graphcodebert}, to assess whether encoder choice matters (Section~\ref{sec:rq1}).
We standardize behavioral features so that no single feature dominates by scale.
We form multimodal representations by concatenating the per-stream vectors, which enables measuring each modality alone and in combination.

\vspace{-3pt}
\subsection{The Contrastive Metric Space}
We compare two models over each representation.
The first is a linear probe, namely $\ell_2$-regularized logistic regression, which serves as a strong and simple baseline and isolates how much signal a representation provides on its own.
The second is a contrastive metric space, in which a two-layer projection head maps representations to a 128-dimensional embedding trained with supervised contrastive loss~\cite{khosla2020supcon}, bringing PRs from the same agent together and pushing those from different agents apart.
At inference, we assign a PR to the agent whose training centroid is most similar to its embedding by cosine similarity.
This metric formulation enables open-set recognition and few-shot enrollment: PRs far from every centroid are flagged as unseen, and new agents are enrolled by adding a centroid from a few labeled PRs, without retraining.
We choose it over a closed-set classifier: though a discriminative classifier over a fixed label set can be comparably accurate on that set (Section~\ref{sec:rq1}), it cannot reject or enroll unseen agents and thus fails the open-world requirement that motivates this study.
We avoid synthetic oversampling: a preliminary experiment with SMOTE~\cite{chawla2002smote} distorted the natural class distribution. Instead, we report results for both a balanced subsample and the natural distribution.
We use the Transformers library~\cite{wolf2020transformers} for the sentence and code encoders, PyTorch~\cite{paszke2019pytorch} to train the contrastive metric space, and scikit-learn~\cite{pedregosa2011scikit} for the linear probe, feature standardization, and cross-validation.

\vspace{-3pt}
\section{Study Design}
\label{sec:design}
\begin{figure*}[t]
\centering
\vspace{-1pt}
\resizebox{.9\textwidth}{!}{
\begin{tikzpicture}[
  font=\footnotesize,
  db/.style={draw, fill=white, cylinder, shape border rotate=90, aspect=0.25,
             minimum width=1.9cm, minimum height=1.0cm, align=center, inner sep=2pt},
  proc/.style={draw, fill=white, rounded corners=2pt, minimum height=1.5cm,
               text width=2.5cm, align=center, inner sep=3pt},
  arr/.style={-{Latex[length=2mm]}, thick},
]
  \node[db] (agent) {AIDev\\agentic \& human\\PRs};
  \node[proc, right=1.2cm of agent] (mod) {\textbf{Four modalities}\\PR text $\cdot$ \\ Commit msgs $\cdot$ \\ Code diffs $\cdot$ Behavioral};
  \node[proc, right=1.2cm of mod, text width=2.8cm,] (enc) {\textbf{Encoders}\\Sentence-BERT (text)\\ Code encoder (diffs)\\ Behavioral features};
  \node[proc, right=1.2cm of enc] (met) {Supervised\\Contrastive\\Metric Space};
  \node[proc, below=0.5cm of met, text width=2.5cm] (base) {Baselines:\\logistic regression, XGBoost};
  \coordinate (mb) at ($(met)!0.5!(base)$);
  \node[proc, right=3.1cm of mb, text width=3.9cm, minimum height=2.9cm, align=left] (rq)
    {\textbf{Research questions}\\[2pt]
     RQ1: Identification\\
     RQ2: Characterization\\
     RQ3: Watermark vs.\ Latent\\
     RQ4: Human-vs-AI Detection\\
     RQ5: Open-world Recognition};
  \begin{scope}[on background layer]
    \draw[fill=white] ($(rq.south west)+(3pt,-3pt)$) rectangle ($(rq.north east)+(3pt,-3pt)$);
    \draw[fill=white] ($(rq.south west)+(6pt,-6pt)$) rectangle ($(rq.north east)+(6pt,-6pt)$);
  \end{scope}
  \draw[arr] (agent.east) -- (mod.west);
  \draw[arr] (mod) -- node[above, font=\scriptsize, align=center] {strip\\markers} (enc);
  \draw[arr] (enc) -- (met);
  \draw[arr] (enc.south) |- (base.west);
  \draw[arr] (met.east) -- node[above, yshift=-23pt, xshift=-8pt, font=\scriptsize, align=center]{$5\times5$ CV\\ F1 / AUC} (rq.west);
  \draw[arr] (base.east) -- (rq.west);
\end{tikzpicture}
}
\vspace{-8pt}
\caption{Overview of our study: the four modalities are stripped of tool markers and encoded, then a supervised-contrastive metric space and classical baselines are evaluated across the five research questions.}
\label{fig:overview}
\vspace{-11pt}
\end{figure*}
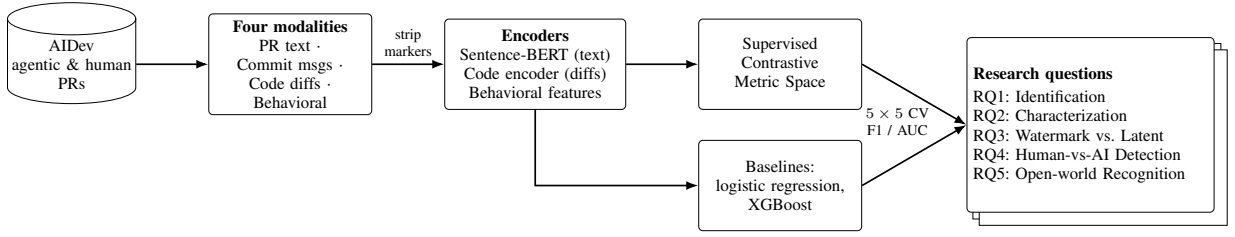

Fig.~\ref{fig:overview} overviews the study, tracing the pipeline from the two data sources, through the four modalities and their encoders, into the contrastive metric space, and out to the five research questions, and situating the dataset, evaluation metrics, and statistical protocol that follow.

\vspace{-3pt}
\subsection{Dataset}
\label{sec:data}
We use \dataset{}~\cite{li2025aidev}, a large dataset of PRs authored by AI coding agents on GitHub together with a sample of human-authored PRs.
We use the curated AIDev-pop subset, which is restricted to repositories with more than 100 stars and enriches each PR with its commits, code changes, and metadata.
After removing PRs with incomplete commit metadata, our set contains \nagentprs{} PRs from five agents, distributed as OpenAI Codex (\NCodexPRs, \PctCodexPRs\%), GitHub Copilot (\NCopilotPRs, \PctCopilotPRs\%), Devin (\NDevinPRs, \PctDevinPRs\%), Cursor (\NCursorPRs, \PctCursorPRs\%), and Claude Code (\NClaudePRs, \PctClaudePRs\%).
This distribution is strongly imbalanced and reflects real-world agent usage rather than a sampling artifact, so we report results under both a balanced subsample and the natural distribution (Section~\ref{sec:rq1}).
The human set contains \nhumanprs{} PRs, sampled by \dataset{} from repositories with more than 500 stars.

\dataset{} does not identify all five agents the same way, which shapes how we treat self-disclosed markers.
GitHub Copilot and Devin open their PRs under dedicated bot accounts, so their label is independent of what the PR says.
Cursor is attributed through its \texttt{cursoragent} committer identity, present in $\CursorCommitterPct\%$ of Cursor PRs, and OpenAI Codex through its platform integration rather than any string in the text.
Claude Code instead commits under the developer's account, so \dataset{} can identify its PRs only by the trailer it appends: a ``\texttt{Generated with Claude Code}'' line and a ``\texttt{Co-Authored-By: Claude}'' line. The ``\texttt{Co-Authored-By}'' line is present in every one of its $\NClaudePRs$ PRs, vs.\ $\ClaudeCoauthorPctOther\%$ of other agentic PRs, so the label and the marker coincide exactly.
Hence, its label is derived from the same self-disclosed string a naive classifier would exploit, so we strip these markers from every representation by default and quantify their contribution separately (Section~\ref{sec:rq3}) rather than allowing them to inflate identification.

A property of \dataset{} shapes our design for AI-versus-human detection.
The dataset provides commit messages and file-level code diffs for agentic PRs, but not for human PRs, for which only PR-level metadata is available.
Obtaining commit messages and diffs for human PRs therefore requires collecting them from the GitHub API, which we do with a per-commit procedure that matches how \dataset{} assembled the agent side so that the two remain comparable.
We collect these for human PRs so that AI-versus-human detection (RQ4) uses the same four modalities as agent identification. We also report the PR-text-only variant, since PR text is the only stream naturally available for both humans and agents, treating stream availability as a first-class dimension of the study (Section~\ref{sec:discussion}).

\subsection{Evaluation Metrics}
We measure identification and AI-versus-human detection with the weighted, macro, and per-class F1 scores, and the rejection of unseen agents with the area under the ROC curve. Enrollment of a new agent is likewise measured by F1, over all agents once the newcomer is added rather than over the newcomer alone, since a metric that ignores the agents already enrolled cannot see what enrolling costs them (Section~\ref{sec:rq5}).
We report the F1 scores under both a balanced subsample of \BalancedCap{} PRs per agent, with the two smallest classes taken in full, and the natural class distribution, since the weighted average alone is dominated by the majority class and can hide minority-agent performance.
Unless a held-out protocol is stated, every estimate comes from five-times-repeated stratified five-fold cross-validation ($25$ train-test estimates), reported with $95\%$ confidence intervals; the RQ-specific procedures (the modality and generalization protocols, the leave-one-agent-out and few-shot enrollment setup, and the AI-versus-human and contamination checks) are described with each research question.

\vspace{-3pt}
\subsection{Behavioral Characterization}
To characterize how agents differ (RQ2), we attribute importance with SHAP~\cite{lundberg2017shap}, which gives an additive and less biased account than gain-based importance, and we complement it with Cliff's delta effect sizes~\cite{romano2006cliff} computed for each feature between an agent and the rest, using the thresholds of negligible, small, medium, and large at $0.147$, $0.33$, and $0.474$.
Effect sizes complement feature importance by showing both the direction and magnitude of each difference, revealing, for example, whether an agent writes longer or shorter commit messages than the others rather than only that message length is discriminative.

\vspace{-3pt}
\subsection{Statistical Adequacy}
For the \NFeatKept-feature behavioral model, even the smallest class (Claude Code, \NClaudePRs{} PRs) yields an events-per-variable ratio above the commonly recommended threshold of ten~\cite{peduzzi1996simulation}, reducing the risk of overfitting.
This heuristic does not apply to the embedding models, so we instead rely on repeated cross-validation, $\ell_2$ regularization for the probe, and report macro and per-class metrics alongside weighted F1.
Model comparisons report the difference in F1, Cliff's delta, and the corrected resampled paired t-test of Nadeau and Bengio~\cite{nadeau1999inference}, declaring significance at $p<0.05$.
We use the corrected test because repeated cross-validation yields dependent folds, and uncorrected tests then underestimate variance. For modality comparisons, we instead report $95\%$ confidence intervals across folds. Rank-based statistics are used where appropriate, namely for the independent agent-versus-rest feature comparisons in RQ2.

\vspace{-2pt}
\section{Evaluation Results}
\label{sec:results}

\subsection{RQ1: How accurately can we identify AI coding agents?}
\label{sec:rq1}

\subsubsection{\textbf{Motivation}}
Attribution is only actionable if it is accurate, so we first ask how accurately the authoring agent of a PR can be identified.
Equally important for deployment is which signal conveys the identity, because the modality an attributor relies on dictates what data it must obtain.
A governance check at PR-open time sees only the title and description, fetching commit messages costs additional API calls, and fetching code diffs is costlier still and, for human PRs, unavailable in existing datasets~\cite{li2025aidev}.
If a cheap modality suffices, attribution becomes far easier to deploy, whereas if the signal exists only in code, every check must fetch and parse diffs.
RQ1 therefore measures both the accuracy of identification and the marginal contribution of each modality.

\subsubsection{\textbf{Methodology}}
We frame identification as a 5-class problem over the five agents and evaluate the linear probe and contrastive metric space (Section~\ref{sec:design}) on each modality and their fusion, under both the balanced subsample and natural class distribution, using stratified 5-fold cross-validation. We report weighted, macro, and per-class F1 with $95\%$ confidence intervals over folds.
To assess whether the code stream weakness is specific to one encoder, we re-encode the code using a general sentence encoder, a code-search encoder, a structure-aware GraphCodeBERT encoder, and code-content features.
Given that a model limited to its training repositories is of little use, we further assess generalization using three held-out protocols: a repository-disjoint split (no repository appears in both train and test, stratified so that every agent appears in each test fold), a leave-one-language-out split over the eight most frequent languages, and a chronological split that trains on the earliest $70\%$ of PRs and tests on the latest $30\%$.

\subsubsection{\textbf{Results}}
Table~\ref{tab:modality} reports identification by modality on the balanced subsample.
The contrastive model matches or exceeds the probe in all settings, with the clearest gains on the text and fusion representations, and the ordering across modalities is the central result: the two text streams together (all text, $\AllTextContr$) nearly match full fusion ($\FusionContr$), the behavioral features alone achieve $\BehavioralContr$, and the code diffs are by far the weakest stream ($\CodeContr$), and adding them to the fusion does not help ($\FusionContr$ drops to $\FusionCodeContr$ for the contrastive model and moves the probe only from $\FusionProbe$ to $\FusionCodeProbe$).
Almost all of the identification signal comes from how a change is described and recorded, not by the code it contains.

\begin{table}[ht]
\centering
\vspace{-6pt}
\caption{Agent identification across modalities (balanced, weighted F1). Probe: logistic regression; Contrastive: supervised contrastive model.}
\vspace{-9pt}
\label{tab:modality}
\resizebox{\columnwidth}{!}{
\begin{tabular}{lrrr}
\toprule
\textbf{Modality} & \textbf{Dimensions} & \textbf{Probe} & \textbf{Contrastive} \\
\midrule
PR text (title + body)          &  768  & \PRTextProbe & \PRTextContr \\
Commit text                     &  384  & \CommitProbe & \CommitContr \\
Code diff (embedding)           &  384  & \CodeProbe & \CodeContr \\
Behavioral features             &   \NFeatKept  & \BehavioralProbe & \BehavioralContr \\
All text (title+body+commit)    & 1,152 & \AllTextProbe & \AllTextContr \\
Fusion (all text + behavioral)  & 1,193 & \FusionProbe & \textbf{\FusionContr} \\
\quad + code diffs              & 1,577 & \FusionCodeProbe & \FusionCodeContr \\
\bottomrule
\end{tabular}
}
\vspace{-5pt}
\end{table}

Under the natural, imbalanced distribution (Table~\ref{tab:imbalance}), weighted F1 achieves $\WFNatural$ (median $\WFNaturalMedian$ over the $25$ folds), driven by the majority class (OpenAI Codex, $\PctCodexPRs\%$ of PRs) being the easiest to identify.
We report this weighted figure as it reflects the natural distribution a deployed attributor faces, and we report macro F1 ($\MFNatural$) alongside it as the class-balanced view, since the two diverge exactly on the minority agents.
The per-class scores locate the difficulty precisely: the three majority agents are near-perfect (Codex $\CodexNaturalTwo$, Copilot $\CopilotNaturalTwo$, Devin $\DevinNaturalTwo$), while the two minority agents lag (Cursor $\CursorNaturalTwo$, Claude Code $\ClaudeNaturalTwo$), which is expected since Claude Code contributes only $\NClaudePRs$ PRs and, as RQ3 shows, most of its separability comes from a self-disclosed trailer that we strip.
The contrastive space helps these classes, increasing macro F1 over the probe, because grouping same-agent PRs sharpens boundaries for small classes that a probe underweights.
For closed-set identification alone the probe is close, within about three points on the fusion, so the contrastive space earns its added complexity not there but through these minority-class gains and through the open-world recognition and enrollment that a closed-set probe cannot provide (RQ5).

\begin{table}[ht]
\centering
\vspace{-6pt}
\caption{Per-agent F1 for balanced \& natural distributions (contrastive model).}
\vspace{-9pt}
\label{tab:imbalance}
\resizebox{\columnwidth}{!}{
\begin{tabular}{p{5cm}rr}
\toprule
\textbf{Agent} & \textbf{Balanced} & \textbf{Natural} \\
\midrule
OpenAI Codex     & \CodexBalanced & \CodexNatural \\
GitHub Copilot   & \CopilotBalanced & \CopilotNatural \\
Devin            & \DevinBalanced & \DevinNatural \\
Cursor           & \CursorBalanced & \CursorNatural \\
Claude Code      & \ClaudeBalanced & \ClaudeNatural \\
\midrule
Weighted average & \WFBalanced & \WFNatural \\
Macro average    & \MFBalanced & \MFNatural \\
\bottomrule
\end{tabular}
}
\vspace{-5pt}
\end{table}

\smallskip\noindent\textbf{Comparison to a baseline.}
To our knowledge, no prior method attributes a PR to a specific coding agent in the open-world setting we target. The closest is the prior closed-set detection~\cite{ghaleb2026fingerprinting}, which assumes a fixed roster of known agents, while the only other agent-fingerprinting work~\cite{zhang2025agentprint} identifies AI coding agents from encrypted network traffic and cannot be applied to a static PR.
As a strong baseline, we therefore run gradient-boosted trees~\cite{chen2016xgboost} on the same 41 features, under the identical repeated cross-validation.
This classical model is strong on closed-set identification: under the natural distribution, XGBoost achieves weighted F1 $\XgbNatW$ and macro F1 $\XgbNatM$.
Neither model dominates, and the two measures disagree.
Under the corrected paired t-test, our contrastive model is significantly higher on weighted F1, the primary measure under the natural distribution ($\OursNatW$ vs. $\XgbNatW$, $p=\PNatW$, Cliff's $\delta=\DeltaNatW$), while XGBoost is significantly higher on macro F1 ($\XgbNatM$ vs. $\OursNatM$, $p=\PNatM$, $\delta=\DeltaNatM$). In contrast, on the balanced sample, the two are indistinguishable on weighted F1 ($p=\PBalW$). Though XGBoost leads in macro F1 by a similar margin, we do not claim this difference because the corrected test is inconclusive: across five random initializations of the projection head the margin persists, but $p$ ranges from $\SeedPMacroLo$ to $\SeedPMacroHi$ (Section~\ref{sec:threats}).
Every difference is at most $\MaxBaselineDiff$ F1, so the tuned feature classifier and our learned space perform similarly, with XGBoost slightly better on rare agents emphasized by macro F1.
We report the split verdict rather than a single winner because the comparison is not our focus: a closed-set feature classifier is a strong identifier, but only an identifier.
Our learned space matches its closed-set accuracy while also providing, from a single model, what a fixed-label tree ensemble cannot: separating agent- from human-authored PRs (RQ4), rejecting and enrolling unseen agents (RQ5), and remaining accurate after self-disclosed markers are stripped (RQ3).

The weakness of the code stream is not an artifact of a single encoder.
Code-only identification stays low under a general sentence encoder ($\CodeMiniLM$), a code-search encoder in the style of CodeBERT~\cite{feng2020codebert} ($\CodeCodeBERT$), a structure-aware encoder pretrained on data-flow, GraphCodeBERT~\cite{guo2021graphcodebert} ($\CodeGraphCodeBERT$), and code-content features, such as comment density, control-flow, and function counts ($\Codehandcraftedcontent$), and adding any of these to the text streams does not improve accuracy (all text moves from $\AllTextNoCode$ to $\AllTextPlusCode$ with the best code encoder).
Even the structure-aware encoder does not help, so the agreement of four independent code representations indicates that the code that AI coding agents generate is inherently less agent-specific than how they describe it.
This inverts the ordering reported for human authorship attribution, where code style dominates~\cite{caliskan2015anonymizing,burrows2007source,abuhamad2018large}, so the code-stylometry methods built to fingerprint human developers do not transfer to AI coding agents, which diverge in how they describe a change rather than in the code, and Fig.~\ref{fig:modality} shows the same modality pattern recurring for AI-versus-human detection (RQ4).

\begin{figure}[t]
\centering
\vspace{-6pt}
\includegraphics[width=\columnwidth]{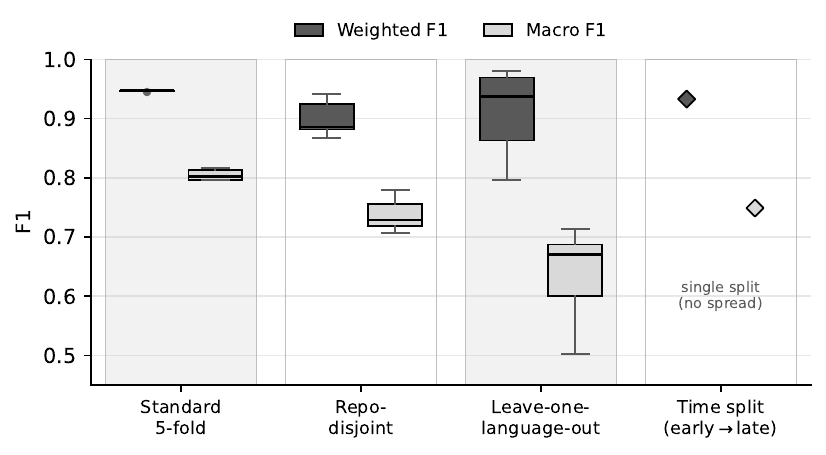}
\vspace{-22pt}
\caption{Robustness of agent identification (all text, contrastive, natural distribution). Boxplots span the folds of each setting; the chronological split is a single split, shown as markers.}
\label{fig:robustness}
\vspace{-10pt}
\end{figure}

\medskip\noindent\textbf{Robustness across repositories, languages, and time.}
A model that identifies AI coding agents only within the repositories it trained on is of little use, so we test whether identification, using the all-text representation, transfers to repositories and programming languages held out of training (Fig.~\ref{fig:robustness}).
Under a repository-disjoint split, where no repository appears in both training and test, weighted F1 falls from $\StdWF$ to $\RepoWF$ and macro F1 from $\StdMF$ to $\RepoMF$ (median $\RepoMFMedian$).
The cost is modest and, more importantly, consistent: the folds sit in a band of about $\SeedRepoBandLo$ to $\SeedRepoBandHi$ macro F1 around that mean, and the mean itself moves by at most $\SeedRepoMeanSpread$ across random initializations (Section~\ref{sec:threats}), so identification transfers to unseen repositories at a stable price rather than degrading unpredictably.
Across programming languages, holding out one language at a time and training on the rest of the eight most frequent languages, weighted F1 stays between $\LangWFLo$ and $\LangWFHi$, but macro F1 varies far more widely, from $\LangMFLo$ on HTML to $\LangMFHi$ on Rust (mean $\LangMFMean$, median $\LangMFMedian$). Each language is a single held-out estimate rather than an average over folds, and the smallest of them are correspondingly unstable, varying by up to $\SeedLoloSpreadHi$ across random initializations vs.\ $\SeedLoloSpreadLo$ for the largest (Section~\ref{sec:threats}). The per-language values therefore locate where transfer is hard rather than measure it precisely, and it is the ordering, stable across initializations, that the finding rests on.
Language is therefore the less forgiving axis of the two: transfer to an unseen repository is cheap and predictable, whereas transfer to an unseen language is cheap on the majority agents and unreliable on the rare ones, which is where the spread lives (Fig.~\ref{fig:robustness}).
That the fingerprint survives an unseen language suggests the signal lies in the PR and commit text, which is largely language-independent, rather than in the code.

Because agents update their templates and underlying models over time, we also split PRs chronologically: the earliest 70\% by submission date for training and the latest 30\% for testing, spanning about seven months (December 2024–July 2025). Weighted F1 drops from $\TimeBaseWF$ to $\TimeWF$ and macro F1 from $\TimeBaseMF$ to $\TimeMF$, indicating that overall identification remains stable while degrading for minority agents. Claude Code, Cursor, and Devin fall to $\TimeClaudeTwo$, $\TimeCursorTwo$, and $\TimeDevinTwo$ on later PRs, mirroring the repository-disjoint results.
Given that the fingerprint is textual, a maintainer can still identify the agent from the PR title and description alone at PR-open time, without fetching or parsing code, though predictions for rare agents should be treated with lower confidence, especially on repositories or time periods unlike the training data.

\begin{tcolorbox}[rqbox]
\textbf{RQ1 Summary.}
\approach{} identifies authoring agents with a weighted F1 of $\WFNaturalTwo$ (macro $\MFNaturalTwo$). PR and commit text alone ($\AllTextContrTwo$) nearly match the full multimodal model ($\WFBalancedTwo$), while code diffs add little under four independent code representations, suggesting agents are distinguished mainly by how they communicate changes rather than the code itself.
\end{tcolorbox}

\vspace{-3pt}
\subsection{RQ2: What behavioral characteristics distinguish AI coding agents?}
\label{sec:rq2}

\subsubsection{\textbf{Motivation}}
Beyond accurate attribution, understanding what distinguishes agents is essential for interpreting and validating the learned fingerprint. These interpretable signatures help maintainers review flagged PRs, help developers understand their tools' behavior, and help researchers track changes in agent behavior, ensuring identification relies on meaningful behavior rather than spurious dataset artifacts.

\subsubsection{\textbf{Methodology}}
We attribute global importance with SHAP~\cite{lundberg2017shap}, which gives an additive and less biased account than the gain-based importance used in prior work~\cite{ghaleb2026fingerprinting}, and we characterize each agent with one-vs-rest Cliff's delta effect sizes~\cite{romano2006cliff}, which add the direction and magnitude that importance alone omits.
Effect size is what separates ``message length distinguishes this agent'' from ``this agent writes shorter messages than the others,'' a distinction that turns out to matter.
We ground the quantitative signatures in representative PRs drawn from each agent.

\subsubsection{\textbf{Results}}
Globally, SHAP importance over the behavioral model ranks the length of the PR description highest, followed by how evenly a change is spread across files and the average commit-message length, so how agents describe and record their changes matters more than what the code looks like.
This mirrors the modality result of RQ1, where the PR-text and commit streams dominate the code stream.

\begin{figure}[ht]
\centering
\vspace{-7pt}
\includegraphics[width=\columnwidth]{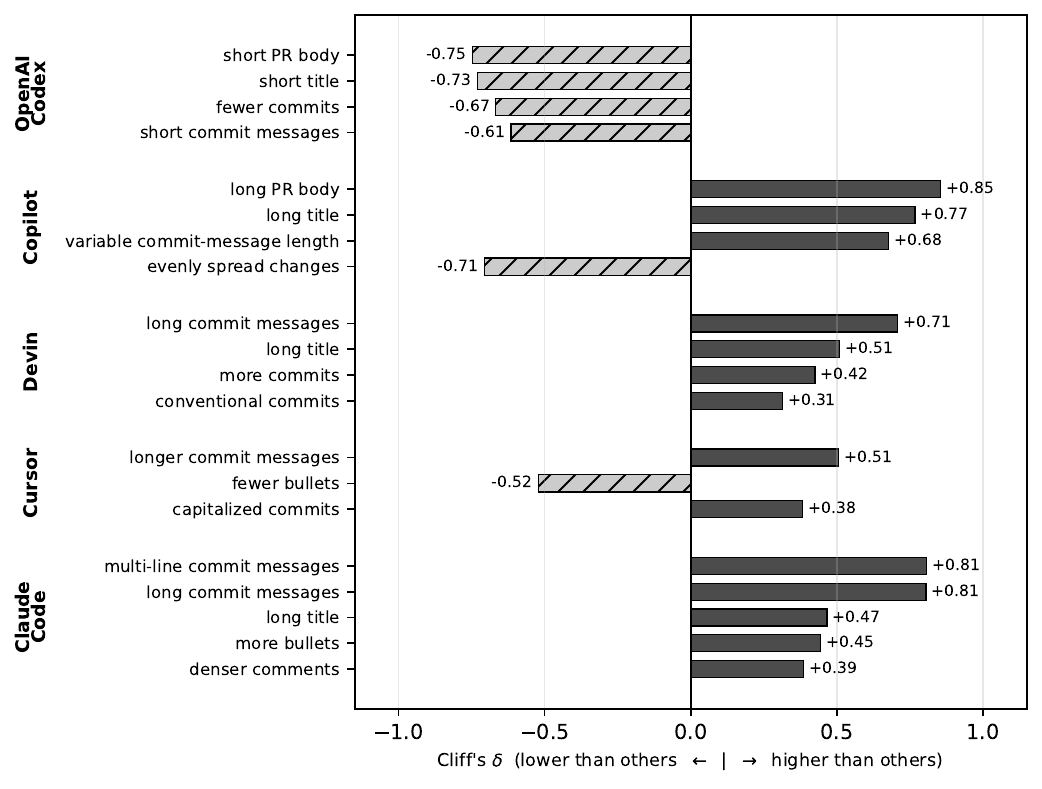}
\vspace{-20pt}
\caption{Behavioral signatures per agent (one-vs-rest, Cliff's delta). Dark bars (right): higher than the other agents; light hatched bars (left): lower.}
\label{fig:signatures}
\end{figure}

Fig.~\ref{fig:signatures} reports, for each agent, the behavioral signals that most distinguish it from the others, with the direction and magnitude given by Cliff's delta.
Reading direction rather than importance alone changes the interpretation.
OpenAI Codex is distinguished by brevity throughout, scoring low on description length ($\delta=\DCodexBody$), title length ($\delta=\DCodexTitle$), number of commits ($\delta=\DCodexCommits$), and commit-message length ($\delta=\DCodexCommitLen$), all large, so it is identified by what it does not write rather than by any distinctive thing it does. This is what an undirected importance score cannot express~\cite{ghaleb2026fingerprinting}, since a feature can be highly discriminative because an agent scores low on it.
Devin, by contrast, writes long commit messages ($\delta=\DDevinCommitLen$) and follows conventional-commit structure more often ($\delta=\DDevinConv$), while Cursor uses fewer bullets ($\delta=\DCursorBullets$) than the others, again opposite to what an undirected score implies.
Copilot writes long descriptions ($\delta=\DCopilotBody$) and distributes changes evenly across files (change-concentration Gini $\delta=\DCopilotGini$), while Claude Code writes the most structured commits of any agent, with the highest multi-line ratio ($\delta=\DClaudeMultiline$), the longest commit messages ($\delta=\DClaudeCommitLen$), and denser inline comments ($\delta=\DClaudeComments$).

Beyond aggregate deltas, several agents show distinctive behaviors (Table~\ref{tab:examples}): $\CopilotInitPlan\%$ of Copilot PRs open with an ``Initial plan'' commit and $\CopilotLinksIssue\%$ link the resolved issue; Devin favors conventional commits ($\DevinConventional\%$). Cursor writes plain, minimally structured descriptions, whereas Claude Code uses much longer commit messages (median $\ClaudeMedCommitChars$ characters after stripping its trailer vs. $\CursorMedCommitChars$ for the next agent), with a summary-plus-bullets format in $\ClaudeSummaryBullets\%$ of PRs. Codex commits are shorter (median $\CodexMedCommitChars$ characters) and, unlike Copilot where $\CopilotSingleLine\%$ of commits are also single-line once trailers are stripped, Codex is uniquely brief on all dimensions.

\begin{table}[t]
\centering
\caption{Distinctive per-agent signatures with representative examples.}
\vspace{-9pt}
\label{tab:examples}
\resizebox{\columnwidth}{!}{
\begin{tabular}{@{}p{1.55cm}p{3.8cm}p{3.5cm}@{}}
\toprule
\textbf{Agent} & \textbf{Distinctive signature} & \textbf{Verbatim example} \\
\midrule
OpenAI Codex & Short on every dimension: very short commit messages (median $\CodexMedCommitChars$ chars), short titles, short bodies, few commits & ``chore(ci): disable web resources CDN'' \\
\addlinespace
GitHub Copilot & Opens with an ``Initial plan'' commit ($\CopilotInitPlan\%$); links the resolved issue ($\CopilotLinksIssue\%$); spreads changes evenly across files & first commit ``Initial plan''; body ``Fixes \#\ldots'' in a \texttt{<details>} block \\
\addlinespace
Devin & Long, conventional commits ($\DevinConventional\%$ conventional) & ``feat: add Open Source Awards section'' with a bulleted body \\
\addlinespace
Cursor & Plain descriptions, little markdown ($\CursorHeaders\%$ use \texttt{\#\#} headers); fewest bullets & ``Disable import button during processing and show loading state'' (no headers) \\
\addlinespace
Claude Code & Long, structured commits: a summary line plus a detailed bulleted body ($\ClaudeSummaryBullets\%$ of PRs, median $\ClaudeMedCommitChars$ chars after stripping, vs.\ $\CursorMedCommitChars$ for the next agent), densest inline comments & ``Fix memory leak: remove global paste listener'' with bullets ``- Store handler reference''; ``- Remove listener in cleanup()'' \\
\bottomrule
\end{tabular}
}
\vspace{-10pt}
\end{table}

\medskip\noindent\textbf{A caution on marker-derived features.}
These signatures are measured on latent text, which makes them behavioral rather than circular. A self-disclosed trailer is more than a string: it takes its own lines and lengthens the message, silently altering features that seem purely behavioral. On raw text, the multi-line commit ratio is $\MultilineCopilotRaw$ for Copilot and $\MultilineDevinRaw$ for Devin, vs. $\MultilineCopilotStripped$ and $\MultilineDevinStripped$ after stripping markers, since a trailer alone makes a one-line commit appear multi-line. An analysis that strips markers from its embeddings but not from its feature extraction will therefore report an agent's trailer as its writing style, and will attribute multi-line commits to whichever agents happen to append one. We accordingly strip before feature extraction, which relocates the multi-line signature from Devin to Claude Code, the one agent whose commits are actually structured. We flag this because the pattern is easy to miss: the feature name gives no hint that a marker feeds it.

These signatures offer concrete, human-readable cues to inspect a suspicious PR without running a model, such as Copilot's ``Initial plan'' opening commit or Claude Code's ``\texttt{Generated with Claude Code}'' trailer.
They also reveal which conventions most expose a tool's identity, providing developers with levers that can be preserved to promote transparency or altered to reduce identifiability.
Two methodological lessons emerge for empirical studies.  Feature importance must be interpreted together with effect direction, since a feature may rank highly because an agent scores low on it. Moreover, self-disclosed markers must be stripped before feature extraction. Otherwise, the resulting features measure disclosure rather than behavior.

\begin{tcolorbox}[rqbox]
\textbf{RQ2 Summary.}
Agents mainly differ in commit-message and PR-description style: Codex is brief on every length signal ($\delta$ from $\DCodexCommitLen$ to $\DCodexBody$), Copilot starts $\CopilotInitPlan\%$ of PRs with an ``Initial plan'' commit, and Claude Code writes the most structured commits ($\delta=\DClaudeMultiline$). Both effect-size direction and marker-free feature extraction are necessary, since using importance alone inverts Codex's signature, and unstripped text makes a trailer mimic a writing style.
\end{tcolorbox}

\subsection{RQ3: Do these fingerprints reflect latent behavior or self-disclosed watermarks?}
\label{sec:rq3}

\subsubsection{\textbf{Motivation}}
Some of what distinguishes an agent is inserted deliberately, such as a PR-template checklist, a conventional-commit prefix, or a trailer like Claude Code's ``\texttt{Generated with Claude Code}'' and ``\texttt{Co-Authored-By}'' lines.
Such markers are effectively a self-applied watermark that an agent, or a developer wishing to hide one, can strip in a single post-processing step.
They are also how \dataset{} identifies some agents in the first place, most acutely Claude Code (Section~\ref{sec:data}), so a model that reads them risks rediscovering the label rather than a behavior, which is why we make the marker-free latent representation the default everywhere.
RQ3 therefore asks the complementary question, namely how much identification the self-disclosed markers actually add on top of latent behavior, and for which agents, which determines both the robustness of detection to evasion and whether the signal reflects genuine behavior.

\subsubsection{\textbf{Methodology}}
We measure the contribution of self-disclosure at two levels.
At the text level, which holds most of the signal, we start from the default latent representation and restore the three marker tiers of Section~\ref{sec:approach} one at a time, using the all-text contrastive model on the natural distribution and reading off per-agent, weighted, and macro F1. The ladder runs from latent, through bare name mentions, then vendor template and tool URLs, to the raw text with trailers restored, so each step isolates one form of self-disclosure. Separating the tiers matters because they are not interchangeable: a bare name is the most direct disclosure imaginable, whereas a trailer is the most conspicuous, and a single ``stripped vs. raw'' contrast would confound the two.
At the behavioral level, we split the 41 features into eight explicit markers, namely the template and structural signals of the description (checklists, bullets, links, code blocks) and the commit-formatting conventions (conventional, multi-line, and capitalized ratios), and the remaining latent signals over code structure, change shape, and timing, and we re-run identification on all features, latent-only, a strict latent set that also drops text-length signals, and markers-only.

\subsubsection{\textbf{Results}}
Table~\ref{tab:watermark} reports the tier ladder.
A latent fingerprint clearly dominates: with all three tiers stripped, all-text identification still achieves weighted F1 $\LadderLatentWF$ (median $\LadderLatentWFMedian$) and macro F1 $\LadderLatentMF$, and restoring every marker raises these only to $\LadderRawWF$ and $\LadderRawMF$. An agent's own name appears somewhere in the raw text, in a trailer, a session URL, or prose, in $\CopilotNameRateRaw\%$ of Copilot PRs, $\DevinNameRateRaw\%$ of Devin PRs, $\CodexNameRateRaw\%$ of Codex PRs, and every Claude Code PR, and yet the whole of this self-disclosure is worth $\DiscloseTotalWF$ weighted F1.
The most informative step is the first. Once trailers and tool URLs are gone, an agent's name still survives in the prose of $\CopilotNameRateProse\%$ of Copilot PRs, $\DevinNameRateProse\%$ of Devin PRs, and $\CodexNameRateProse\%$ of Codex PRs, and restoring those names improves weighted F1 by $\NamesDeltaWF$. The names are not empty of signal: on the same text, a five-rule regex that only checks for each agent's name achieves a weighted F1 of $\NameRegexProse$, versus $\NameRegexFloor$ after stripping the names, so the names alone add $\NameRegexGainSigned$ to a rule with no other features.
Hence, the name is redundant, not uninformative, since latent style already identifies these agents, so naming the author adds little beyond what the model infers from the PR's writing. This is the best response to the concern that the model merely reads a watermark.
The remaining gain concentrates where the label itself is a marker. Vendor template adds $\BoilerDeltaWF$ and trailers a further $\TrailerDeltaWF$, and the per-agent view shows that trailers are worth at most $\TrailerCostMaxMajor$ to Codex and Copilot but lift Claude Code from $\LadderBoilerClaude$ to $\LadderRawClaude$.
Claude Code is exactly the agent whose \dataset{} label is derived from its trailer, so the latent $\LadderLatentClaude$ is the non-circular estimate of how identifiable it is from behavior alone, and everything above it measures its self-disclosure.
Behavioral analysis confirms the independent-feature conclusion: over the \NFeatKept{} features, stripping the eight explicit markers only slightly lowers weighted F1 from $\FeatAllWF$ to $\FeatLatentWF$, the latent features are more discriminative than the markers alone ($\FeatLatentWF$ vs.\ $\FeatMarkersWF$), and a strict latent set that also drops text-length signals still achieves $\FeatStrictWF$, far above chance.
Across both analyses, the pattern holds: Codex, Copilot, and Devin remain highly identifiable from latent behavior, whereas Cursor and Claude Code depend more on presentation, so removing explicit markers alone does not prevent attribution for most agents.

\begin{table}[t]
\centering
\vspace{-1pt}
\caption{Per-agent F1 as self-disclosure is restored one tier at a time, from latent (our default) to raw; all-text contrastive, natural distribution.}
\vspace{-9pt}
\label{tab:watermark}
\resizebox{\columnwidth}{!}{
\begin{tabular}{p{2.6cm}rrrr}
\toprule
 & \textbf{Latent} & \textbf{$+$Names} & \textbf{$+$Boiler-} & \textbf{$+$Trailers} \\
\textbf{Agent} & \textbf{(default)} & & \textbf{plate} & \textbf{(raw)} \\
\midrule
OpenAI Codex     & \LadderLatentCodex & \LadderNamesCodex & \LadderBoilerCodex & \LadderRawCodex \\
GitHub Copilot   & \LadderLatentCopilot & \LadderNamesCopilot & \LadderBoilerCopilot & \LadderRawCopilot \\
Devin            & \LadderLatentDevin & \LadderNamesDevin & \LadderBoilerDevin & \LadderRawDevin \\
Cursor           & \LadderLatentCursor & \LadderNamesCursor & \LadderBoilerCursor & \LadderRawCursor \\
Claude Code      & \LadderLatentClaude & \LadderNamesClaude & \LadderBoilerClaude & \LadderRawClaude \\
\midrule
Weighted average & \LadderLatentWF & \LadderNamesWF & \LadderBoilerWF & \LadderRawWF \\
Macro average    & \LadderLatentMF & \LadderNamesMF & \LadderBoilerMF & \LadderRawMF \\
\bottomrule
\end{tabular}
}
\vspace{-7pt}
\end{table}

Trailers and templates alone are insufficient for policy enforcement, since they can be removed by users who wish to avoid attribution. Robust detection relies on the latent behavioral fingerprint that persists after marker removal.
However, the reverse holds for developers of AI coding agents, since a tool that wants to signal AI authorship transparently should keep its markers, whereas a tool designed to blend into human workflows would have to alter not only its trailers but its latent commit and description style, which is harder to do without degrading usefulness.
Detection is nonetheless not evasion-proof, since a determined adversary who mimics human latent style could lower accuracy further, for instance by instructing the agent through a system prompt or a project file such as \texttt{CLAUDE.md} to write in a plain, human-like style, which is lower-effort than modifying the tool itself, a limitation addressed in Section~\ref{sec:threats}.

\begin{tcolorbox}[rqbox]
\textbf{RQ3 Summary.}
Behavioral fingerprints persist after stripping trailers, tool URLs, vendor template, and every mention of an agent's own name, with weighted F1 remaining at $\LadderLatentWFTwo$ vs.\ $\LadderRawWFTwo$ on raw text. Restoring the removed markers provides little benefit overall, except for Claude Code ($\LadderLatentClaudeTwo \rightarrow \LadderRawClaudeTwo$), whose dataset label is derived mainly from its trailer.
\end{tcolorbox}

\vspace{-3pt}
\subsection{RQ4: To what extent can we separate AI-authored from human-authored PRs?}
\label{sec:rq4}

\subsubsection{\textbf{Motivation}}
Identifying which agent authored a PR assumes the PR is agent-written, but the motivating scenario is the opposite: a developer runs an agent locally and submits its output under a human account, so the account appears human while the true author is an agent.
Distinguishing agents from humans is the key deployment scenario for governance and for dataset validity, because datasets that label ``human'' vs.\ ``AI'' by submitter identity~\cite{li2025aidev} systematically mislabel undisclosed AI cases, and such label noise can mislead downstream empirical work~\cite{ghaleb2019noise}.
RQ4 addresses whether the two can be separated and how far existing ``human'' labels are contaminated.

\subsubsection{\textbf{Methodology}}
Because \dataset{} provides no commit or diff tables for human PRs, we collected them from the GitHub API using the same per-commit procedure that assembled the agent side (Section~\ref{sec:data}), successfully retrieving commit histories for $\NHumanWithCommits$ of the \nhumanprs{} human PRs ($\PctHumanWithCommits\%$), so that both classes share the four modalities.
The multimodal comparison uses these $\NHumanWithCommits$ PRs, whereas the contamination scan below uses all \nhumanprs{}, since it relies only on PR text.
We train the contrastive model to separate the two classes on balanced samples and measure each modality and the fusion.
To establish that the separation reflects who wrote the PRs rather than where they come from, we apply three progressively stricter confound controls: matching agent repositories to the human sample by star count, restricting to the repositories that contain both classes, and a repository-disjoint split within those shared repositories.
Finally, we estimate label contamination two ways, by counting explicit AI-authorship markers in the human PRs and by scoring each human PR by its out-of-fold probability of being agent-authored, and we manually inspect the highest-scored cases.

\subsubsection{\textbf{Results}}
\begin{figure}[ht]
\centering
\vspace{-3pt}
\includegraphics[width=\columnwidth]{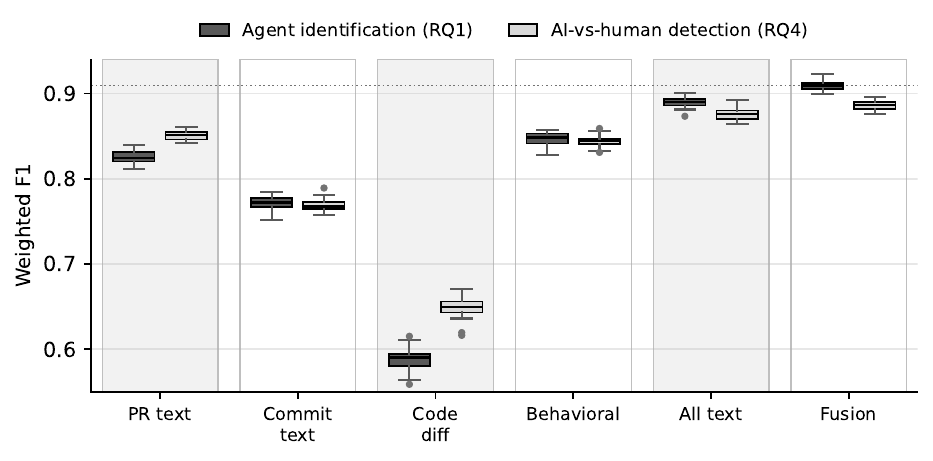}
\vspace{-22pt}
\caption{Weighted F1 by modality for AI coding agent identification (RQ1) and AI-versus-human detection (RQ4); balanced sets, contrastive model. Boxplots span the $25$ folds of the repeated CV; the dotted line marks the best fusion median.}
\label{fig:modality}
\vspace{-11pt}
\end{figure}

With the collected data, detection can use the same four modalities as agent identification (Fig.~\ref{fig:modality}).
From PR text alone, the setting available without any collection, the contrastive model separates AI- from human-authored PRs at a balanced F1 of $\HvAPRText$.
Adding commit text raises this to $\HvAAllText$ (all text), and the full fusion achieves $\HvAFusion$ (median $\HvAFusionMedian$), while the code stream alone is much weaker ($\HvACode$). As in agent identification, detection relies on how a change is described rather than on the code, a pattern that now holds on a second, independent task.
Because \dataset{} samples human PRs from more highly starred repositories than agentic PRs, part of this separation may reflect repository differences rather than authorship.
Restricting agentic PRs to repositories with more than 500 stars, which matches the human sample, the multimodal model still separates the two at $\HvAStarMatchedFusion$.
The cheaper PR-text-only detector remains robust under a stricter control, namely restricting to the \NSharedRepos{} repositories containing both classes and using a repository-disjoint split, where it holds at $\HvAWithinRepo$.
The separation therefore reflects authorship rather than repository population.

We quantify label contamination using two complementary analyses.
First, independent of any classifier, $\ContamFloorN$ of the \nhumanprs{} human PRs ($\ContamFloorPct\%$) contain explicit evidence of AI authorship, including tool-named co-authorship trailers, ``Generated with'' lines, session URLs, vendor template, or \texttt{anthropic}, \texttt{cursoragent}, or \texttt{devin} signatures. These PRs are thus agent-assisted despite being labeled as human by submitter identity. We count only AI-specific markers: generic ``Co-authored-by'' and ``Signed-off-by'' trailers, common in human practice ($\GenericTrailerPct\%$ of human PRs), would otherwise inflate the estimate by about threefold.
Second, we score each human PR by its out-of-fold probability of being agent-authored under the repository-matched classifier (which never sees markers, since we strip them), finding that $\ContamFiveN$ ($\ContamFivePct\%$) exceed $0.5$ probability and $\ContamNineN$ ($\ContamNinePct\%$) exceed $0.9$.
This is circular (i.e., human labels are both the training signal and the object under audit) so we treat it as an upper bound. The marker floor nonetheless validates it sharply: the explicit-marker rate rises with classifier confidence, from $\ContamFloorPct\%$ at base to $\ContamFiveMarker\%$ above $0.5$, $\ContamSevenMarker\%$ above $0.7$, and $\ContamNineMarker\%$ above $0.9$, a $\ContamNineEnrich\times$ enrichment. Thus a marker-blind model trained only on latent style concentrates disclosed agent authorship at the top of its ranking, which would not occur if the flag were noise.
A manual reading of the $\AuditN$ highest-scored human PRs, all scored above $\AuditPMin$, supports this: $\AuditDisclosed$ carry explicit AI disclosure, a rate consistent with the $\ContamNineMarker\%$ above, and $\AuditTemplate$ use a structured PR template (a \texttt{<details>} block, a ``\#\# Summary'' section, or a task checklist) together with conventional-commit style typical of AI agents, with every disclosing PR among those $30$. The remaining $\AuditInconclusive$ are inconclusive, because an undisclosed agentic PR leaves no clear marker.
These findings show that a dataset separating human from AI by submitter identity has measurable label noise, at least $\ContamFloorPctZero\%$ by disclosure and likely more, that a text-based check can detect.

Undisclosed AI authorship is detectable from the PR text alone, without any code access, which is the information a maintainer or a dataset curator actually has at hand, so a lightweight text-only check can flag likely agent contributions at open time or during dataset construction.
The contamination we surface, at least $\ContamFloorPctZero\%$ by explicit disclosure and plausibly more by latent style, is a warning to researchers, namely that labels derived from submitter identity in datasets such as \dataset{}~\cite{li2025aidev} should be validated with a content-based check or treated as noisy, since building conclusions about ``human'' code on a set that mixes in agent-assisted PRs risks the measurement error that label noise is known to induce~\cite{ghaleb2019noise}.

\begin{tcolorbox}[rqbox]
\textbf{RQ4 Summary.}
AI- and human-authored PRs separate at a balanced F1 of $\HvAFusion$ ($\HvAPRText$ from PR text alone), remaining at $\HvAWithinRepo$ under repository-disjoint evaluation within shared repositories. At least $\ContamFloorPctZero\%$ of the dataset's ``human'' PRs contain explicit AI-authorship markers, and a marker-blind classifier concentrates them $\ContamNineEnrich\times$ at high confidence, indicating measurable label contamination that a text-only check can surface.
\end{tcolorbox}

\vspace{-13pt}
\subsection{RQ5: To what extent can we detect unseen agents and enroll them from few examples?}
\label{sec:rq5}

\vspace{-1pt}
\subsubsection{\textbf{Motivation}}
The five coding agents we study represent a fast-changing landscape where new agents appear frequently.
A closed-set classifier must assign every PR to an existing label, so it cannot detect PRs from unseen agents, and retraining for each new tool is costly.
Our target deployment is open-world: detecting when a PR matches none of the known agents (open-set recognition~\cite{scheirer2013openset}) instead of mislabeling it as the nearest one, and enrolling a newly released agent from a few labeled PRs without retraining the embedding (few-shot enrollment~\cite{snell2017prototypical}).
This RQ addresses whether the learned metric space, unlike a closed-set classifier, supports both, and how much of each ability comes from learning the space rather than from the representation it is learned over.

\subsubsection{\textbf{Methodology}}
We use a leave-one-agent-out protocol: we train the metric space on four AI coding agents, hold out the fifth as unseen, and score each test PR by its maximum similarity to the four known-agent centroids, where a PR from the held-out agent should score low.
We report the Area Under the Receiver Operating Characteristic (AUC) of separating known from unknown PRs, averaged over the five choices of held-out agent.
For enrollment, we add the held-out agent as a prototype centroid formed from ten labeled PRs~\cite{snell2017prototypical}, without retraining, and then classify both the held-out agent's remaining PRs and the known agents' test PRs over all five prototypes.
Scoring the known agents alongside the new one is essential. Reporting only the new agent's recall, as is common, measures how many of its PRs the new prototype captures but not what capturing them costs the agents already enrolled, so a prototype that draws PRs indiscriminately scores well on a metric that never looks at the damage. We therefore report the F1 of the enrolled agent and the macro F1 over all five agents after enrollment, consistent with the F1 reporting used throughout the paper.
To isolate the contribution of learning, we compare with a non-learned baseline identical in every other respect: the same representation, held-out agents, folds, seeds, ten-shot enrollment protocol, and scoring, with only the trained projection head replaced by the identity. Matching the representation is essential, since comparing learned and unlearned models across different modalities would conflate representation choice with the benefit of learning.

\subsubsection{\textbf{Results}}
Table~\ref{tab:openworld} reports both capabilities under the leave-one-agent-out protocol.
With the full multimodal representation, the learned metric space separates known from unknown PRs with a mean AUC of $\OpenFusionL$ across the five held-out agents, vs.\ $\OpenFusionN$ for the matched non-learned baseline, so learning the space is worth $\OpenGain$ AUC over embedding the same features without it.
This mean should be read with its spread, since it averages over five agents rather than over folds and two of them are unusually easy to flag: Devin achieves $\OpenDevin$ and Cursor $\OpenCursor$, while the remaining three sit close together at $\OpenRestLo$ to $\OpenRestHi$, so the median is $\OpenMedian$ and describes the typical agent better than the mean does.
The ordering is interpretable rather than noise. Devin, the most distinctive agent, stays easy to flag even with its templates and session URLs stripped, whereas at the other end two agents are hard to flag for different reasons. OpenAI Codex is hard despite contributing $\PctCodexPRs\%$ of the training PRs, because a space fitted mostly to Codex treats a held-out Codex PR as familiar rather than foreign, a difficulty that is structural rather than an artifact of the sample. Claude Code scores marginally lower still, but on only $\NClaudePRs$ PRs its AUC is the noisiest of the five, so we read its position as small-sample variance rather than a property of the agent.
Learning gains increase with representation richness, from $\OpenGainPRText$ on PR text to $\OpenGain$ on fusion, and the ordering across representations mirrors RQ1, with the text streams carrying most of the signal.

Enrollment shows the same advantage: adding the held-out agent from ten labeled PRs yields a macro F1 of $\EnrollFusionL$ over the resulting five agents, vs.\ $\EnrollFusionN$ without a learned space, and learning wins on every representation.
However, when the new agent is considered in isolation, since the non-learned metric achieves an F1 of $\NewAgentN$ on the enrolled agent vs.\ $\NewAgentL$ for the learned space, again on every representation.
The two views are consistent: an unlearned metric captures more of the newcomer's PRs but partly by reassigning PRs from existing agents, trading accuracy for new-agent coverage, whereas the learned space fits the newcomer at a lower cost to the rest. Because a deployed attributor must identify all agents simultaneously, macro F1 is the relevant metric, even though new-agent recall alone suggests the opposite conclusion.

\begin{table}[t]
\centering
\vspace{-2pt}
\caption{Leave-one-agent-out open-world recognition. AUC for unseen-agent detection and macro F1 after enrolling the unseen agent from ten labeled PRs.}
\vspace{-9pt}
\label{tab:openworld}
\resizebox{\columnwidth}{!}{
\begin{tabular}{lrrrr}
\toprule
 & \multicolumn{2}{c}{\textbf{Open-set AUC}} & \multicolumn{2}{c}{\textbf{Macro F1 (enrolled)}} \\
\cmidrule(lr){2-3}\cmidrule(lr){4-5}
\textbf{Representation} & \textbf{Learned} & \textbf{Non-learned} & \textbf{Learned} & \textbf{Non-learned} \\
\midrule
PR text (title + body)        & \OpenPRTextL & \OpenPRTextN & \EnrollPRTextL & \EnrollPRTextN \\
Commit text                   & \OpenCommitL & \OpenCommitN & \EnrollCommitL & \EnrollCommitN \\
All text                      & \OpenAllTextL & \OpenAllTextN & \EnrollAllTextL & \EnrollAllTextN \\
Fusion                        & \textbf{\OpenFusionL} & \OpenFusionN & \textbf{\EnrollFusionL} & \EnrollFusionN \\
\bottomrule
\end{tabular}
}
\vspace{-5pt}
\end{table}

Open-world recognition prepares a detector for the steady arrival of new agents, since a governance system can flag a PR as ``an agent, but not one we know'' rather than silently mislabel it as the nearest known agent, which is the failure mode a closed-set classifier cannot avoid, and can add a newly released tool from ten labeled PRs the day it appears.
An AUC of $\OpenFusionL$ supports triage rather than adjudication, so unknown-agent flags are best used to route PRs for review rather than to decide authorship on their own, and the margin over a non-learned metric ($\OpenGain$ AUC, $\EnrollGain$ macro F1) sets a realistic expectation of what the learned geometry provides.
Enrollment results warrant a caution similar to the RQ2 effect-size warning: evaluating a newly enrolled class only by its own recall favors prototypes that capture its PRs, regardless of what they take from existing classes, which can reverse which method appears superior.

\begin{tcolorbox}[rqbox]
\textbf{RQ5 Summary.}
The learned metric space detects unseen agents (AUC $\OpenFusionL$ vs.\ $\OpenFusionN$ without metric learning) and enrolls a new agent from ten labeled PRs with macro F1 $\EnrollFusionL$ across all five agents (vs.\ $\EnrollFusionN$). Though the non-learned baseline has higher recall on the new agent alone ($\NewAgentN$ vs.\ $\NewAgentL$), it degrades performance on previously enrolled agents, making single-class evaluation misleading.
\end{tcolorbox}

\vspace{-3pt}
\section{Discussion and Implications}
\label{sec:discussion}

\subsection{What drives the fingerprint}
Two findings explain the attribution results across the research questions.
First, text is the dominant signal: PR descriptions and commit messages nearly match the full multimodal model for both agent identification (RQ1) and AI-versus-human detection (RQ4), while code is the weakest modality across four representations. This inverts human authorship attribution, where code style is the primary signal~\cite{caliskan2015anonymizing,abuhamad2018large,burrows2007source}, and suggests that agents, optimized to produce conventional code, are distinguished more by how they communicate changes than by the code they generate.
Second, the fingerprint is behavioral rather than a watermark: it persists after stripping self-disclosed markers (RQ3), generalizes across held-out repositories and languages (RQ1), and is stable enough to detect and enroll unseen agents (RQ5), which mainly differ in communication style (RQ2).

\vspace{-3pt}
\subsection{Habits versus watermarks}
Attribution rests on distributed behavioral patterns, but the per-agent signatures of RQ2 split into two types that deployments should handle differently. This creates a second choice axis alongside access cost. Cost concerns what data an attributor must collect. This axis concerns how much data remains if an agent stops declaring itself, which is crucial for governance, where disclosure is voluntary and a trailer can be removed with one configuration line.
Copilot's ``Initial plan'' opening commit is a workflow habit: it identifies the agent without being intended to, and it survives stripping. Claude Code's trailer is the opposite, a deliberate self-disclosure that is also how \dataset{} defines the label.
The tier ladder of RQ3 quantifies the difference per agent, which is uneven to a degree the aggregate hides. Removing every marker costs $\MarkerCostCodex$ F1 for OpenAI Codex and $\MarkerCostCopilot$ for GitHub Copilot, but $\MarkerCostClaude$ for Claude Code, a spread of about thirty times (Table~\ref{tab:watermark}). Hence, exposure to evasion depends less on how the attributor is built than on which agent it targets: Codex and Copilot are identifiable from habits they do not control, so an evader gains almost nothing by stripping its trailers, whereas Claude Code falls to $\LadderLatentClaude$, the only agent a single post-processing step moves from reliably identified to barely identified.
This warns against reading a per-agent F1 as a fixed property, since the $\LadderRawClaude$ we measure for Claude Code with its markers present describes how well it is identified while it cooperates. An attributor built on watermarks is as fragile as the watermarks, and a study that measures them is correspondingly circular.

\vspace{-3pt}
\subsection{Accuracy, access cost, and deployment choices}
\begin{figure}[ht]
    \centering
    \vspace{-7pt}
    \includegraphics[width=\columnwidth]{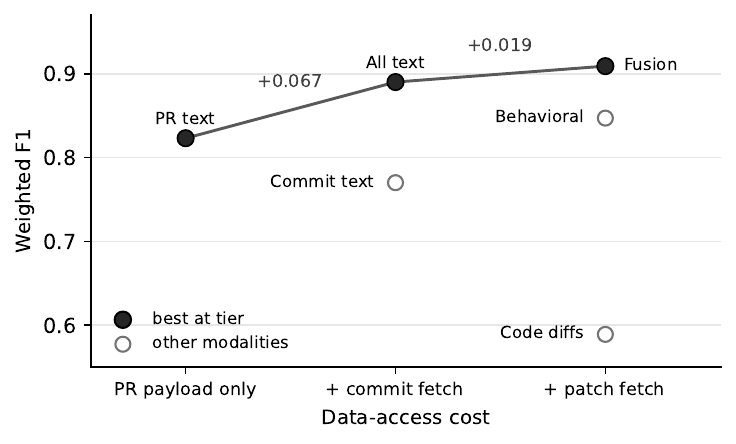}
    \vspace{-20pt}
    \caption{Accuracy versus data-access cost. Filled markers linked by line: the best modality reachable at each tier; hollow: the others at that tier.}
    \vspace{-10pt}
    \label{fig:tradeoff}
\end{figure}

The modalities vary widely in acquisition cost: PR text comes with the PR itself, commit messages need commit history, and code diffs require retrieving and processing patches. Compute is not the binding constraint, since encoding a PR on CPU takes about $6$ ms for its text and $16$ ms for all four streams, so deployment is constrained by data retrieval, not computation.
Considering that cost (Fig.~\ref{fig:tradeoff}), returns diminish quickly: commit text adds $\TradeoffCommitGain$ weighted F1 over PR text, patch access adds only $\TradeoffPatchGain$ more, and code diffs, the most expensive stream to obtain, are the least accurate modality. Effective attribution therefore does not require code access. When only the PR is available, PR text alone is a strong practical choice. With commit access, combining PR text and commit messages provides the best accuracy-to-cost trade-off and is our default recommendation. Full fusion is justified only when feature-level explanations or marginal accuracy gains warrant the additional cost of processing diffs.

\vspace{-3pt}
\subsection{Implications}
\noindent\textbf{For repository maintainers and platforms.}
Attribution from PR text alone enables a lightweight check at PR-open time (RQ1, RQ4), backed by per-agent signatures a reviewer can inspect by eye (RQ2).
This differs from bot detection, which flags accounts that are meant to be recognizable through metadata and message templates~\cite{wessel2018power,golzadeh2022accuracy}, since agents use ordinary developer accounts, so the signal must be behavioral, and RQ3 shows it persists even after removing the self-disclosed markers a naive evader would target.
Such a check complements code-level detectors of AI-generated code~\cite{tian2023chatgpt,Nguyen2024gptsniffer,bisztray2025know} by operating at the PR level and, per RQ1 and RQ4, without needing the code at all.

\smallskip\noindent\textbf{For AI coding agent developers.}
The signatures of RQ2 are levers for a deliberate choice about disclosure: an agent that wants to be transparent can preserve its distinctive markers, while one that aims to blend into human workflows would have to alter not only its trailers but its latent description and commit style (RQ3), which is hard to change without degrading the usefulness of its messages.

\smallskip\noindent\textbf{For researchers.}
Datasets that distinguish human from AI by submitter identity are unreliable, as PR text alone separates the two and shows that at least $5\%$ of \dataset{}'s human PRs self-report agent assistance (RQ4).
Thus, labels based on submitter identity~\cite{li2025aidev} should be validated with content checking or treated as noisy, since such contamination can bias studies of ``AI'' versus ``human'' code, as label noise can distort empirical results~\cite{ghaleb2019noise}.
Three measurement choices can determine conclusions rather than refine them, and each is easy to get wrong by default. Self-disclosed markers must include tool URLs, vendor template, and plain tool names, not just trailers, or most agents still appear as latent text (RQ3). Stripping must precede feature extraction, not just before embedding, or a trailer is treated as writing style and misassigns a behavioral signature to the wrong agent (RQ2). An enrolled class must be scored together with existing classes, or a metric that ignores that cost reverses which method seems better (RQ5). Each of these choices inverted a result we would otherwise have reported.
More broadly, the open-world results (RQ5) support building attributors that expect new agents rather than a fixed roster, keeping longitudinal studies valid as tools change.

\vspace{-3pt}
\section{Threats to Validity}
\label{sec:threats}

\noindent\textbf{Construct validity.}
Some \dataset{} labels are derived from the same text used for attribution, most notably Claude Code, whose label comes from its commit trailer (Section~\ref{sec:data}). Because every one of the $\NClaudePRs$ Claude Code PRs includes a ``\texttt{Co-Authored-By: Claude}'' line, identifying the agent from unstripped text would be circular. We therefore strip all tool-identity markers by default and assess their contribution separately in RQ3, making the latent result (RQ3, $\LadderLatentClaude$) the only non-circular estimate of Claude Code identifiability. The other four agents are labeled independently through bot accounts, committer identity, or platform integration, so textual markers correlate with but do not restate their labels. A few PRs contain markers from multiple agents (e.g., \CursorPRsWithClaudeTrailer{} Cursor PRs include a Claude Code trailer), indicating multi-agent workflows, but these markers are stripped before attribution. Because our split between explicit markers and latent signals (RQ3) is also a design choice, we disclose the eight stylistic markers used.

What counts as a self-disclosed marker is itself a design choice, and a narrow definition silently inflates results: stripping only commit trailers leaves vendor template and session URLs that name the tool, so the PRs of three of the five agents still disclose their author (RQ3). We therefore strip in three cumulative tiers (Section~\ref{sec:approach}), reducing tool mentions to at most $\MaxNameResidual\%$ of PRs for four of the five agents. The exception is Claude Code at $\ClaudeNameResidual\%$, where the residual is not disclosure but subject matter: the surviving strings are code identifiers and branch names that embed the tool's name (e.g., \texttt{createClaudeSession}), which tier~3 does not remove because it deletes standalone name tokens rather than substrings of identifiers. Since Claude Code is also the agent with the lowest latent F1 ($\LadderLatentClaude$), this residual is not what drives its identification.

The repository-disjoint split is stratified by agent and grouped by repository. Though grouping alone is common, given that OpenAI Codex supplies $\PctCodexPRs\%$ of PRs, it can fill an entire fold: an unstratified grouping yields a test fold of \DegenNTest{} Codex PRs containing no other agent, whose macro F1 averages four absent classes as zero and reads $\DegenMF$ while its weighted F1 is $\DegenWF$. Such a fold measures the composition of the split rather than the transfer of the model, and it moves the reported mean below every fold that is actually informative. Stratifying by agent keeps repositories disjoint while guaranteeing that each test fold can score every class.
The code stream is also limited to the first $4{,}000$ added characters, mean-pooled, and encoded without fine-tuning or AST information, which may understate its value. Yet all four independent code representations, including the hand-crafted code-content features ($\Codehandcraftedcontent$) that are not affected by truncation or pooling, still perform poorly, suggesting code is less agent-specific than text.

\smallskip\noindent\textbf{Internal validity.}
The AI-versus-human analysis may be confounded by repository population because \dataset{} samples human PRs from repositories with more than 500 stars and agentic PRs from repositories with more than 100 stars. We mitigate this by matching repositories by star count, where the multimodal detector still achieves $\HvAStarMatchedFusion$, and by evaluating repository-disjoint splits within the \NSharedRepos{} repositories containing both classes, where the PR-text detector still achieves $\HvAWithinRepo$, indicating that attribution reflects authorship rather than repository population.

Human commit and diff data were collected separately from the agent data, introducing a potential collection bias. We replicate \dataset{}'s per-commit collection procedure (Section~\ref{sec:data}), and the PR-text-only detector, which relies solely on \dataset{}'s original tables for both classes, already achieves $\HvAPRText$, showing that the main result does not depend on the commit-collection pipeline. The classifier-based contamination estimate is reported as an upper bound because it is inherently circular, whereas the marker-based estimate provides an independent $5\%$ lower bound. To account for class imbalance, we report macro, weighted, and per-class F1, avoid synthetic oversampling, and report 95\% confidence intervals from repeated cross-validation.

\smallskip\noindent\textbf{External validity.}
The study is limited to GitHub, a single dataset snapshot, and five AI coding agents, so results may not generalize to other platforms, private repositories, or more agents. Within \dataset{}, attribution transfers to held-out repositories and programming languages, with a modest macro F1 drop across repositories ($\StdMF \rightarrow \RepoMF$) and larger variation across languages ($\LangMFLo$ to $\LangMFHi$), both concentrated in the smallest agent classes.
The reported fingerprints also reflect the current behavior of these agents rather than permanent identities. Models, templates, prompting conventions, repository instructions (e.g., \texttt{AGENTS.md} and \texttt{CLAUDE.md}), custom system prompts, and project conventions can all alter an agent's style. Consistent with this, macro F1 drops from $\TimeBaseMF$ to $\TimeMF$ over the seven-month collection period, and Claude Code falls to $\TimeClaude$ on later PRs (Section~\ref{sec:rq1}), indicating measurable drift even within the observed window. Some of the strongest signatures, such as Copilot's ``Initial plan'' commit and fixed commit trailers, are therefore likely version-specific rather than intrinsic. Deployed attribution models should be periodically revalidated, and even though latent behavioral fingerprints remain after removing explicit markers, prompt engineering or manual editing may still reduce attribution accuracy.

\smallskip\noindent\textbf{Conclusion validity.}
We report Cliff's delta for effect sizes, following recommendations for empirical software engineering~\cite{arcuri2011practical}, and evaluate both balanced and natural class distributions. Estimates come from five-times-repeated five-fold cross-validation ($25$ estimates) except where a held-out protocol is required by the question itself, namely the repository-disjoint, leave-one-language-out, and chronological splits of RQ1 and the leave-one-agent-out protocol of RQ5, each of which is compared with a baseline computed under that same protocol rather than with a figure taken from a different one. We compare models to the baseline using the corrected resampled paired t-test~\cite{nadeau1999inference} and interpret modality differences conservatively using 95\% confidence intervals, avoiding over-interpretation of small gaps.

The projection head is trained stochastically, so each estimate using it is a single draw, and the fold confidence intervals reflect data-split variance, not initialization. Repeating the fragile estimates across five random initializations, with splits and data held fixed, leaves the aggregate results unchanged: weighted F1 under the natural distribution, the tier ladder, and the repository-disjoint mean each shift by at most $\SeedRepoMeanSpread$. Three small-sample estimates are less stable. Claude Code's per-class F1 varies by $\SeedClaudeSpread$, so its third decimal, and that of the latent estimate in Section~\ref{sec:rq3}, should not be read as precise. The repository-disjoint folds span about $\SeedRepoBandLo$ to $\SeedRepoBandHi$. Leave-one-language-out, where each language is a single held-out estimate, varies by $\SeedLoloSpreadLo$ for \SeedLoloSmallLang{} but by $\SeedLoloSpreadHi$ for the much smaller \SeedLoloBigLang{}. The per-language ordering is nonetheless stable, with HTML the hardest and Rust the easiest in every initialization.

We report the mean across folds, since confidence intervals and paired tests are based on it, while also providing the median for each reported result. Across the $25$-fold distributions (modality, marker-tier, and AI-versus-human), the mean and median agree within $\MeanMedianMaxGap$. The choice matters most for the two results that average over a few units: open-set recognition on five agents, where two cases lift the mean AUC ($\OpenFusionL$) above the median ($\OpenMedian$), and leave-one-language-out on eight languages, where two small languages drop the mean ($\LangMFMean$) below the median ($\LangMFMedian$).

\vspace{-3pt}
\section{Related Work}
\label{sec:related}

\textbf{Empirical studies of AI coding agents and agentic PRs.}
A growing body of work characterizes agentic PRs without attributing them.
Li et al.~\cite{li2025aidev} introduced \dataset{}, nearly one million agentic PRs from five agents across over one hundred thousand repositories, and showed that agentic and human PRs differ in merge rates and task distributions, while Siddiq et al.~\cite{siddiq2026security} analyzed security-related agentic PRs in the same data and identified signals linked to rejection.
Closest to our behavioral characterization, two studies examine how the same five agents communicate on \dataset{}: Watanabe et al.~\cite{watanabe2026communicate} characterize per-agent description style and its effect on review, and Gong et al.~\cite{gong2026message} find that agent descriptions often claim changes the diff does not contain, echoing our finding that agent identity shows more in how a change is described than in the code.
Other work reports that agent PRs contain stronger commit messages but weaker summaries~\cite{pham2026agentic} and higher description-to-diff similarity~\cite{ogenrwot2026how}, and studies contribution outcomes such as PR acceptance across agents~\cite{pinna2026comparing}, reasons PRs fail to merge~\cite{ehsani2026where}, how often agents modify CI/CD configurations and with what success~\cite{ghaleb2026cicd}, and adoption trends on GitHub~\cite{robbes2026agentic}.
Unlike these studies, our work attributes which agent or human authored a change, and quantifies the label contamination that account-based labels introduce.

\smallskip\noindent\textbf{Code authorship attribution and detection of automated contributions.}
Attributing code to its author has traditionally targeted human programmers from stylistic characteristics: byte- and token-level n-gram profiles~\cite{burrows2007source,frantzeskou2006effective}, lexical, structural, and syntactic features for de-anonymization~\cite{caliskan2015anonymizing}, language-oblivious neural models~\cite{abuhamad2018large}, AST- and path-based representations~\cite{zhang2019novel,bogomolov2021authorship}, and general-purpose code encoders such as CodeBERT~\cite{feng2020codebert} and GraphCodeBERT~\cite{guo2021graphcodebert}. Most recently, language models are evaluated for attributing human-written code~\cite{dipongkor2025reassessing}.
Recent surveys~\cite{kalgutkar2019code,horvath2026bridging} report a focus on closed-world attribution based on code style, with open-world settings and behavioral signals relatively underexplored.
Other studies detect whether code is AI-generated: Tian et al.~\cite{tian2023chatgpt} use perplexity and burstiness, GPTSniffer~\cite{Nguyen2024gptsniffer} fine-tunes CodeBERT on ChatGPT snippets, and Bisztray et al.~\cite{bisztray2025know} attribute code to generating models through stylometry. These methods operate on isolated code snippets, whereas we attribute complete PRs and find that textual signals are more informative than the code itself.

Closest to our study, Ghaleb~\cite{ghaleb2026fingerprinting} distinguishes the five agents in a closed-set setting from commit, PR-structure, and code features. We extend this to multimodal learned representations, open-world recognition and few-shot enrollment, AI-versus-human attribution, robustness after stripping self-disclosed markers, and modality trade-offs. AgentPrint~\cite{zhang2025agentprint} attributes agents from encrypted network traffic, achieving an F1 of $0.866$ for closed-set identification. We instead attribute agents from public PR artifacts alone and support unseen-agent recognition, few-shot enrollment, and AI-versus-human attribution.
Finally, bot detection identifies automated accounts from metadata, activity patterns, and message templates~\cite{wessel2018power,golzadeh2022accuracy}; unlike those openly-labeled bots, human-mediated coding agents use ordinary developer accounts, so attribution must rely on behavioral traces in the artifacts rather than account-level signals.

\vspace{-3pt}
\section{Conclusion}
\label{sec:conclusion}

We proposed \approach{}, a multimodal, learned, open-world approach for attributing AI coding agents from pull requests (PRs), and evaluated it on \nagentprs{} agent-authored and \nhumanprs{} human-authored PRs from \dataset{}.
\approach{} learns a supervised contrastive metric space over PR text, commit messages, code diffs, and behavioral features, identifying the authoring agent with a weighted F1 of $\WFNaturalTwo$ (macro F1 $\MFNaturalTwo$).
The results show that AI coding agents are distinguished primarily by how they communicate changes rather than by the code they generate, with PR and commit text nearly matching the full multimodal model while code contributes little.
Behavioral fingerprints remain detectable after stripping self-disclosed markers, enabling attribution that does not rely on explicit disclosures.
The same representation separates AI- from human-authored PRs, reveals measurable contamination in human-labeled datasets, and supports recognition and few-shot enrollment of previously unseen agents without retraining.
Together, these findings establish that AI coding agents leave robust behavioral fingerprints and provide a practical foundation for attribution from publicly available software artifacts.

\medskip\noindent\textbf{Future work.}
Future work should extend the study to additional agents and newer releases, evaluate stronger structure-aware code encoders (e.g., AST- and program-graph-based models), and conduct larger manual audits of flagged PRs to refine contamination estimates.
Improving few-shot enrollment remains a key priority, as it is the weakest capability evaluated. Neither episodic prototypical training nor augmenting the learned embedding with raw features improved over the supervised contrastive space, suggesting that enrollment requires representations designed for novel classes.
Future work should also test robustness to adversarial evasion, such as marker removal and style manipulation, and quantify how repository-specific instruction files (e.g., \texttt{CLAUDE.md}, \texttt{AGENTS.md}) affect behavioral fingerprints and attribution accuracy.

\section*{Acknowledgement}
This work is funded by the Natural Sciences and Engineering Research Council of Canada (NSERC): RGPIN-2025-05897.

\balance
\bibliography{paper}
\bibliographystyle{IEEEtran}

\end{document}